\documentclass{aa}
\input psfig.tex
\usepackage{graphicx}
\usepackage{natbib}
\bibpunct{(}{)}{;}{a}{}{,} 
\usepackage{amssymb}
\usepackage{latexsym}
\bibpunct{(}{)}{;}{a}{}{,}

\def\smallskip{\vskip 4pt}

\begin{document}

\title{ Shells of dust around AGB stars: effects
on the integrated spectrum
       of Single Stellar  Populations}


     \author{Lorenzo Piovan, Rosaria Tantalo \and Cesare Chiosi}

     \offprints{L. Piovan }

     \institute{Department of Astronomy, University of Padova,
                Vicolo dell'Osservatorio 2, 35122 Padova, Italy \\
                \email{piovan@pd.astro.it;
                 \, tantalo@pd.astro.it; \, chiosi@pd.astro.it}
     }

     \date{Received: January, 2003; Revised: April, 2003; Accepted: June, 2003}

\abstract{ In this paper we present models for Single Stellar
Populations (SSPs)  of
intermediate and old ages where dust enshrouded Asymptotic Giant
Branch (AGB) stars
are introduced. As long known AGB stars are  surrounded by dust-rich
shells of matter caused by their own stellar wind, which absorb the
radiation coming from the central object
and re-emit it in the far infrared (IR).
To this aim, particular care is devoted to
follow the evolution of the AGB stars throughout the quiet and
thermally pulsing
regimes, to evaluate the effect of self contamination in the outermost
 layers by the
third dredge-up mechanism, to follow the transition from
oxygen-rich to carbon-rich objects (as appropriate to their
initial mass and chemical composition), and finally to estimate
the efficiency of mass-loss by stellar winds, all aspects that
concur to the formation and properties of the dusty shells around.
In addition to this, accurate physical models of the dusty shells
are presented in which the re-processing of radiation from the
central stars is calculated by solving the radiative transfer
equations in presence of dust particles of different chemical
composition. The resulting spectral energy distribution (SED) is
examined to show how important features, like the $10 \mu m$
$Si-O$ stretching mode feature and the $11 \mu m$ $SiC$ feature,
evolve with time. The SEDs are then convolved with the IRAS
filters to obtain the flux in various pass-bands, i.e. 12, 25 and
60 $\mu m$, for individual AGB stars of different, mass, chemical
composition, and age. The comparison is made by means of SSPs
along which AGB stars of the same age but different initial masses
are located. This allows us to explore the whole range of masses
and ages spanned by AGB stars.
 The theoretical results  are compared
 to the observational data for selected groups of stars. The same
is made for the J,H,K, L pass-bands of the Johnson system.
Finally, from the integrated SEDs  of the SSPs, we derive the
integrated Johnson J,H,K, L magnitudes and colors to  be compared
to infrared data for star clusters of the Magellanic Clouds. In
general good agreement with the data is possible if the effects of
the circumstellar shells of dust are taken into account.
\keywords{stars: structure -- stars: AGB -- stars: mass-loss --
stars: spectra -- stars: dust -- spectra: infrared} }

\titlerunning{Shells of dust around AGB stars }
\authorrunning{L.Piovan, R. Tantalo \&\ C. Chiosi}
\maketitle

\section{Introduction}

The first all-sky survey at far-IR wavelengths carried out in 1983
by the Infrared Astronomical Satellite (IRAS) has opened a new era
of modern infrared astronomy. Thousands of galaxies were detected
to emit most of their light in the IR. After IRAS, a long  series
of observations were started to explore the IR Universe. In the
COBE all-sky maps a bright isotropic background (CIRB) was
discovered in the far-IR/sub-mm spectral regions, whose origin is
nowadays attributed to  the integrated emission by dust in
primeval galaxies absorbing and scattering the stellar light and
returning it at long wavelengths \citep{Pug96}. For the first
time, the Infrared Space Observatory (ISO)  surveyed and detected
distant galaxies in the mid- and far-IR, allowing us to know in
detail their emission in those spectral regions \citep{Elb99}.

It became soon  clear that precious information on the star
formation history (SFH) of galaxies and the Universe as a whole is
hidden in the UV-optical  and IR ranges of the spectral energy
distribution, thus spurring cross-correlated studies of the two
regions. Many wide-field and all-sky surveys are currently running
or have been just completed, e.g. the Galaxy Evolution Explorer
[GALEX] \citep{Mar97} and the Sloan Digital Sky Survey [SDSS]
\citep{Yor00} in the UV-optical range; the Two Micron All-Sky
Survey [2MASS] \citep{Skr97} and the Deep Near-Infrared Survey of
the Southern Sky [DENIS] \citep{Epc97} in the near infrared.
Combined with other astronomical databases, they  provide a huge
amount of UV-optical and near IR photometric data for millions of
galaxies. In particular, the observational information in the
infrared will further grow with the new surveys SIRTF (Space
Infrared Telescope Facility, see
\textit{http://sirtf.caltech.edu/SSC/}), UKIDSS (the successor of
2MASS, see \textit{http://www.ukidss.org/}) and with the advent of
the Next Generation Space Telescope (NGST/JWST).

This wealth of data must be accompanied by a
continuous upgrade of the basic theoretical tools to fully exploit
any information on the physical properties of the observed
objects. In particular, to understand  the role played by dust as
strong IR emitter is mandatory.  The key instruments for
photometric studies are the SSPs,  the building blocks of the
assemblies of stars of different complexity going from star
clusters to galaxies, and the fundamentals of population synthesis
techniques. An SSP is defined as a coeval, chemically homogeneous
assembly of stars all contributing to build up the integrated
SED in a way proportional
to the duration of their evolutionary phase, luminosity and
relative number (the luminosity function), in other words
according to the precepts of Fuel Consumption Theorem
\citep{Tinsley80,Renbuz83}.

Let us shortly summarize here the various steps by which the
theoretical SED of an SSP is derived, (a) given an age, the
corresponding isochrone in the HRD is divided in elemental
intervals small enough to assure that the luminosity, gravity, and
$T_{eff}$ in them are nearly constant (the isochrone is
approximated to series of virtual stars, for each of which we know
the spectrum); (b) in each elemental interval the star mass spans
a suitable range $\Delta M$ fixed by the evolutionary rate,
therefore  the number of stars  per elemental interval is
proportional to the integral of the initial mass function (IMF)
over the range $\Delta M$ (the differential luminosity function);
(c) finally, the contribution to the flux at each wavelength of
the spectrum by each elemental interval is weighed on  the number
of stars in it and their luminosity. To conclude, basic
ingredients of the SSP SEDs  are the isochrones and their path in
HRD (in turn functions of the initial chemical composition), the
IMF, and a library of stellar spectra for different values of
$T_{eff}$, gravity, and chemical composition.

With a few exceptions to be mentioned later, the light emitted by
SSPs is modeled neglecting the presence of dust around their stars
\citep[e.g.]{Ber94,Tantalo98a,Gir02}. However, there are at least
two situations in which dust is present: the very initial stages
when stars are born embedded in molecular clouds, and the late
stages of AGB stars when dusty envelopes are formed. In this study
we will concentrate on the AGB stars leaving to a forthcoming
paper the case of young stars.

The dust shells surrounding AGB stars are the result of mass-loss
by stellar winds and the complex structure and evolution of these
stars during the thermally pulsing AGB phase (TP-AGB). They trap
the radiation coming from the central AGB star and re-emit it in
the far IR \citep[see][for a classical review of this
topic]{Hab96}. The contribution to the IR radiation by the dusty
shells of AGB stars is particularly relevant because AGB stars are
luminous objects able to significantly affect the integrated SED
of star clusters and galaxies.

A great deal of studies have been devoted to model the circumstellar
shells around AGB stars and to understand in detail important
properties of the shell structure, such as the composition of the dust
grains, the dust-to-gas ratio, the expansion velocity of the matter,
and the mass-loss rates \citep[cf.][]{Hab96}. However, only an handful
of studies have tried to include the effect of the dust shells
surrounding AGB stars on the integrated SEDs of SSPs: the pioneer
paper by \citet{Bgs98} and the recent series of three articles by
\citet{MouI02,MouII02} and \citet{MouIII02},

There are several causes hampering detailed studies of the
subject:

(i) First of all, theoretical spectra of oxygen-rich (O-stars) and
carbon-rich (C-stars) AGB stars, surrounded by shells of matter and
even of simple AGB stars are not available. Only empirical spectral
are to our disposal. This is an obvious difficulty for any fully
theoretical analysis of the problem. The spectra of AGB stars of any
type are particularly difficult to obtain because of the many
parameters entering the problem, i.e.  the spectrum of the inner
stellar source, the optical depth (key parameter of the radiative
transfer problem to be solved to follow the propagation of radiation
throughout the dusty shell), and the properties of dust grains in the
shell. These latter depend on the optical depth in such a way that we
have different compositions for different optical depths and of course
passing from O-stars to C-stars.

(ii) Second, for long time SSPs have been developed to interpret
observational data in the optical range of the spectrum. Only recently
the wealthy of data in the IR have spurred several groups to develop
suitable SSPs extending to the far infrared. The study of
\citet{Bgs98}, despite severe limitations such as the use of giant
spectra to model the radiative transport in O-stars and C-stars, the
lack of the transition between O-stars and C-stars, and the crude
treatment of the dust composition, has opened the way to a new
generation of SSPs in the far IR. \citet{MouI02} and
\citet{MouIII02} have improved the situation including
appropriate models for the AGB phase, the transition between O-stars
and C-stars, and an empirical library of spectra of long period
variables (LPV) as input for the circumstellar shells.  It is
important to remind here that while \citet{Bgs98} made use of purely
theoretical spectra, \citet{MouI02} and \citet{MouIII02} have adopted
an empirical library of spectra of AGB stars, so that part of their
results is implicit in this initial assumption.

Aim of this study is to go further along the theoretical line of
work and to generate modern integrated SEDs of SSPs that: (i)
extend to the far IR; (ii) include more updated models of AGB
stars; (iii) allow for the metallicity dependence and finally (iv)
include an accurate modelling of the shells of dust surrounding
these stars, thus improving upon the emission in the IR spectral
region. We will follow a purely theoretical approach as in
\citet{Bgs98} and will try to include a detailed treatment of the
circumstellar dusty shells and of the transition from O- to
C-stars. It is clear that the theoretical spectra of M giant stars
are not as good as the empirical ones (they can indeed be applied
only to the case of unobscured O- and C-stars). However, we will
insist on the purely theoretical approach in order to establish
how far current theory can go in interpreting and reproducing the
observational data.

This paper is organized in the following way. In
Sect.~\ref{par:Dusty} we present the model for the envelope of AGB
stars, describe the assumptions for the radiative transfer
problem, and calculate an expression for the optical depth, the
key parameter of the radiative transfer equations, and its
dependence on basic stellar parameters such as the star mass $M$,
radius $R$, effective temperature $T_{eff}$, luminosity $L$,
pulsation period $P$, and metal content $Z$. In
Sect.~\ref{par:massloss} we briefly summarize the prescription we
have adopted for the mass-loss rates of AGB stars. In
Sect.~\ref{par:dustoptical}, we discuss in detail the optical
properties of dusty AGB shells. As a matter of fact, the dust
absorption coefficients are different for O-stars and C-stars, and
even for the same type of stars, different optical depths may
imply different properties. In Sect.~\ref{par:parameters} we
present our choice for the parameters governing the dusty shells
of AGB stars, i.e, the temperature on the inner boundary, the
density profile across the shells, and the type and mixture of
grains. In Sect.~\ref{uncerta} we try to discuss and evaluate the
uncertainty affecting the models of the dust shells and how this
would reflect onto the SED of the SSPs with particular attention
to the IR range. In Sect.~\ref{theo_SSP} we define the
monochromatic flux emitted by a SSP, summarize the assumption for
the initial mass function and the rates of mass-loss from stars
that we have adopted, report on the libraries of stellar models,
isochrones and stellar spectra with aid of which we have
calculated our SSPs, and finally discuss in detail the problem of
the transition between O-stars and C-stars AGB stars and how this
is incorporated in our SSPs.  In particular we describe how we
have included the recent models by \citet{Mar99} into the SSP
models by \citet{Tantalo98a}.  In Sect.~\ref{par:SSPs} we present
the SEDs of our new SSPs that take the effect of dusty AGB shells
into account and compare our results with those for the standard
SSPs \citep{Tantalo98a}. In Sect.~\ref{par:IRcolors} we describe
the integrated IRAS far IR colors predicted by our models and
compare them with the observational data for a sample of AGB stars
(Mira, Semi-Regular Variables, Long Period Variables, OH/IR stars
and C-stars). We also derive the integrated near IR colors of the
SSPs and compare them with the data for a selected sample of young
star clusters of the Magellanic Clouds. In addition to this, we
examine the age dependence of the integrated near IR colors of
SSPs from different sources, i.e.
\citet{Ber94,Tantalo98a,Gir02,MouI02,MouII02} and this study.
Finally, some concluding  remarks are drawn in
Sect.~\ref{summary}.

\section{Modelling a dusty envelope}\label{par:Dusty}

The problem of radiative transfer in the dusty shells surrounding
the mass losing stars (AGB stars in particular) has been addressed
by many authors \citep[see ][ and references therein]{Hab96}. The
most complete formulation of the problem couples the radiative
transfer and hydro-dynamical equations for the motion of the two
interacting fluids, gas and dust, and takes  the interdependence
between gas, dust and radiation pressure into account.

Since our goal is to simply include the effects of the AGB star dusty
shells on the spectra of SSPs, we will limit ourselves to consider the
problem of radiative transfer in the shell and will leave
hydrodynamics aside. To this aim, we need to link the fundamental
parameters of a star, i.e.  mass $M$, radius $R$, luminosity $L$,
pulsational period $P$, and metal abundance $Z$ to the parameters that
characterize the shell of matter and that are relevant to the solution
of the radiative transfer across it.

The radiative transfer equation is

\begin{equation}
\frac{dI_{\lambda}\left(l\right)}{dl}=
  k_{\lambda}\left(l\right)
    \left[S_{\lambda}\left(l\right)-I_{\lambda}\left(l\right)\right]
\label{fiore}
\end{equation}

\noindent
where

\begin{enumerate}
\item[-] $k_{\lambda}\left(l\right)$ is the overall extinction
coefficient at the wavelength $\lambda$ given by the sum of the
absorption and scattering coefficients,
$k_{\lambda}\left(l\right)=k_{a\lambda}\left(l\right)+k_{s\lambda
}\left(l\right)$;

\item[-] $I_{\lambda}\left(l\right)$ is the intensity of
the radiation field;

\item[-] $S_{\lambda}\left(l\right)$ is the ``\emph{source
function}'', given by the ratio
$\varepsilon_{\lambda}\left(l\right)/k_{\lambda}\left(l\right)$,
with $\varepsilon_{\lambda}\left(l\right)$ the emission
coefficient.
\end{enumerate}

As demonstrated by \citet{Row80} and \citet{Ive95,Ive97},
only two dimensional scales are involved in  the radiative
transfer equation because $I_{\lambda}$ and $S_{\lambda}$ have the
dimension of an intensity, whereas $k_{\lambda}$ and $l$ have the
dimension of a length ($k_{\lambda}$ is the inverse of a length).
Therefore, any physical quantity related to the radiative transfer
problem can be expressed as  functions of these two scales.

Defining the dimensionless element of optical depth along the ray path
$dl$ as $d\tau_{\lambda}=k_{\lambda}\left(l\right)dl$, we can write
eq. (\ref{fiore}) as

\begin{equation}
\frac{dI_{\lambda}\left(l\right)}{d\tau_{\lambda}}
=S_{\lambda}\left(l\right)-I_{\lambda}\left(l\right)
\end{equation}

Starting from these simple considerations, \citet{Ive97} have proved
that the radiative transfer equation satisfies the property of scale
invariance: the physical dimension of any system can be increased and
decreased in an arbitrary way without affecting the radiative
properties, as long as optical depths and spatial variation of the
opacity remain the same. Two systems with different dimensions and
absorption coefficients, but with the same total optical depths and
auto-similar distributions of opacities and ``\emph{source
functions}'' will produce the same intensity of the radiation field.

\citet{Row80}  first  applied the radiative transfer scale
invariance to the IR emission of a central source surrounded by a dust
shell. Subsequently, \citet{Ive97} presented a general formulation of
the problem in arbitrary geometry and distribution of dust, and
studied in detail the case of a spherical shell of dust heated up by a
central source. They also pointed out that the concept of scale
invariance is particularly useful when the absorption coefficient does
not depend on the radiation intensity. Unfortunately, as a consequence
of this, the analysis by \citet{Ive97} cannot be applied to emission
or photoionization lines (where the absorption coefficient can depend
on the intensity of the radiation field via its effects on the level
populations) but only to the continuum of the radiation coming from
dust heated by a central source. Last point to note is that when the
shell is optically thin over all wavelengths $\lambda$, the
approximation $\tau _{\lambda}\sim 0$ may be used thus obtaining a
quick analytical solution of the problem. In contrast, when the shell
is optically thick the problem must be solved numerically.

In the following we apply the method, the results and the
numerical code DUSTY by \citet{Ive97} to study the dust shells
surrounding AGB stars. The spherical symmetry approximation is
adopted for the sake of simplicity. The key parameter we
need to solve the radiative transfer problem and to calculate the
emerging flux is the optical depth $\mathcal{T}_{\lambda}$ of
the shell which is defined as follows

\begin{equation}
\mathcal{T}_{\lambda} = \int_{S}d\tau _{\lambda}\left(r\right) =
\int_{S}k_{\lambda} \left(r \right) \rho
\left( r \right) dr
\label{taut}
\end{equation}

\noindent
where $k_{\lambda}$ is the overall extinction coefficient per mass
unit, $\rho$ is the matter density.  Both depend on the radial
distance $r$ from the central source. The integral is evaluated over
the thickness $S$ of the shell. In the case of a dusty shell
eq. (\ref{taut}) becomes

\begin{equation}
\mathcal{T}_{\lambda} =
\int_{S} k_{\lambda} \left(r \right) \rho_{d} \left( r \right) dr
\label{tau}
\end{equation}

\noindent
where $\rho_{d} \left( r \right)$ is the dust density.

The equation of mass conservation is given by

\begin{equation}
\frac{dM \left(r \right)}{dt} =
4 \pi r^{2} \rho \left( r \right) v \left( r \right)
\end{equation}

\noindent
The matter density $\rho \left( r \right) $ is linked to the dust
density $\rho_{d} \left( r \right) $ by the relation $\rho_{d} \left(
r \right) = \rho \left( r \right) \delta $ where $\delta$ is the
dust-to-gas ratio. Substituting the matter density $\rho$ with the
dust density $\rho_d$ we obtain

\begin{equation}
\frac{dM \left(r \right)}{dt} =
    \frac{4 \pi r^{2} \rho_{d} \left( r \right) v \left( r
\right)}{\delta}
\end{equation}

\noindent
and then

\begin{equation}
\rho_{d} \left( r \right) = \frac{\dot{M} \left( r\right) \delta}{
4 \pi r^{2} v \left( r \right)}
\end{equation}

\noindent
With aid of the above relation we recast the optical depth of
eq. (\ref{tau}) as

\begin{equation}
\mathcal{T}_{\lambda} = \int_{S} k_{\lambda} \left(r \right)
\frac{\dot{M} \left(r \right) \delta}{ 4 \pi r^{2} v \left( r
\right)}  dr
\label{taubis}
\end{equation}

In reality,  the dusty shell will extend between an internal and
external radius, $r_{in}$ and  $r_{out}$ respectively.
Assuming $\delta$ to be constant across the shell, we get

\begin{equation}
\mathcal{T}_{\lambda} = \frac{\delta}{4 \pi}\int_{r_{in}}^{r_{out}}
k_{\lambda} \left(r \right) \frac{\dot{M} \left( r\right)}{r^{2} v
\left( r \right)}  dr
\label{taubis}
\end{equation}

To proceed further, the mass-loss rate $\dot{M(r)}$, velocity
$v(r)$, and extinction coefficient $k_{\lambda}(r)$ and their
radial dependence  must be specified. To a first approximation we
assume that at any given time the rate of mass-loss and the
velocity are constant with $r$, together with physical properties
of the dust. Therefore the absorption coefficient for unit mass
$k_{\lambda}$ does not depend on the radial coordinate $r$. With
these simplifications we have

\begin{equation}
\mathcal{T}_{\lambda} = \frac{\delta \dot{M} k_{\lambda}}{4 \pi
v}\int_{r_{in}}^{r_{out}} \frac{1}{r^{2}}  dr
\label{tauter}
\end{equation}

\noindent
and upon integration

\begin{equation}
\mathcal{T}_{\lambda} = \frac{\delta \dot{M} k_{\lambda}}{4 \pi v}
\left(\frac{1} {r_{in}} - \frac{1}{r_{out}} \right)
\label{tauter}
\end{equation}

\noindent
Since $r_{out}$ is usually much larger than $r_{in}$, the relation
above can be safely approximated to

\begin{equation}
\mathcal{T}_{\lambda} \simeq
\frac{\delta \dot{M} k_{\lambda}}{4 \pi v} \frac{1}{r_{in}}
\label{tauter}
\end{equation}

Now we need to connect the physical quantities defining the
optical depth of the dusty shell with the typical parameters of AGB
stars, e.g. mass $M$, radius $R$, effective temperature $T_{eff}$,
period $P$ of  pulsation, luminosity $L$, and metal content $Z$.
Assuming  the shell to be optically thick to IR radiation (a good
approximation in case of high mass-loss rates),  the inner radius
of the shell can be derived from the equality

\begin{equation}
L = 4 \pi R_{*}^{2} \sigma T_{eff}^{4} = 4 \pi r_{in}^{2} \sigma T_{d}^{4}
\label{equality}
\end{equation}

\noindent
where $R_{*}$ is the stellar radius and $T_{d}$ is the temperature of
dust condensation at $r_{in}$. From this relation we get

\begin{equation}
r_{in} =
\left( \frac{L_{\odot}}{4 \pi \sigma T_{d}^{4}} \right)^{\frac{1}{2}}
\left(\frac{L}{L_{\odot}} \right)^{\frac{1}{2}}
\label{inner_shell}
\end{equation}

\noindent The uncertainty arising from using the relation
(\ref{equality}) can transferred to $T_d$ which is not firmly
established. Literature values range from 800 to 1500
\citep{Row82,Dav90,Suh99,Suh00,Lor00,Lor01,Suh02}. Finally,
adopting $T_{d}$= 1000 K (see Sect.~\ref{par:parameters} below) we
obtain

\begin{equation}
r_{in} =
2.37 \cdot 10^{12} \left(\frac{L}{L_{\odot}} \right)^{\frac{1}{2}}
         \qquad {\rm cm}
\end{equation}

The extinction coefficient for unit mass $k_{\lambda }$ is in general
given by

\begin{displaymath}
k_{ \lambda} = \frac{\sum_{i} n_{g_{i}} \sigma_{g_{i}}}{\rho_{d}}
\end{displaymath}

\noindent
where the summation is extended over all types of grain in the
mixture, $\sigma_{g_{i}}$ is the cross section for the i-th type of
dust grain and $n_{g_{i}}$ is the number of grains of i-th type for
unit volume. Denoting with $m_{g_{i}}$ the mass of the i-th type of
dust grains, and introducing the mass abundance $\chi_{i} = n_{g_{i}}
m_{g_{i}} / \rho_{d}$ of i-th type of grain we finally obtain

\begin{equation}
k_{\lambda} =\sum_{i} \chi_{i} \frac
{ \sigma_{g_{i}}}{  m_{g_{i} }}
\end{equation}

\noindent The optical depth $\mathcal{T}_{\lambda}$ of the dust is
a function of $k_{\lambda}$, and $k_{\lambda}$ in turn is a
function of $\mathcal{T}_{\lambda}$ via the mass abundance
$\chi_{i}= \chi_{i} \left( \mathcal{T}_{\lambda} \right)$. In our
models, the mass abundances $\chi_{i}$ may change for two reasons.
Firstly we are dealing with O- and/or C-stars.  Secondly for the
same type of star (either O- or C-stars) the opacities and the
relative abundances of the grains can change with the optical
depth (see Sect.~\ref{par:dustoptical}). Since $k_{\lambda}$ and
$\mathcal{T}_{\lambda}$ are each other interwoven, an iterative
procedure is required \footnote{The radiative transfer code
"DUSTY" refers the optical depth $\mathcal{T}_{\lambda}$ at a
wavelength of reference.  Similarly to \citet{Suh99,Suh00,Suh02},
we assume as reference wavelength $\lambda = 10 \mu m$. This
implies that the term $k_{\lambda}$ in the optical depth equation
has to be evaluated at this wavelength, once the dust composition
is fixed.}.

We also need the cross sections of the radiation-dust
interactions. The cross section for a single dimension of the
grains $a$ is $\sigma_{g_{i}}\left(a\right) = \pi a^{2}Q_{ext}
\left(i\right)$, where $Q_{ext} \left(i\right)$ are the extinction
coefficients.  The total cross section $\sigma_{g_{i}}$ can be
obtained integrating the cross section for a single dimension over
a given distribution of the dimensions of the grains (see
Sect.~\ref{par:parameters} for a discussion about the choice we
have made for the distribution). The absorption and scattering
coefficients, $Q_{abs}\left(i\right)$ and $Q_{sca}\left(i\right)$
respectively, can be calculated at any given wavelength with the
aid of the Mie theory \footnote{The Mie theory, developed in 1908
by Gustav Mie to understand the colors generated by gold particles
suspended in water, provides a formal solution for the interaction
of spherical and homogeneous small particles with the
electromagnetic radiation \citep{Boh83}.}. The parameters to be
specified are: the optical constants of the material composing the
dust grains, the dimension and shape of the grains that are
usually approximated to a sphere. Basing on this, the absorption
and scattering coefficients, $Q_{abs}$ and $Q_{sca}$, can be
derived from the complex dielectric function, expressed by the
real $n\left(i\right)$ and the imaginary part $k\left(i\right)$ of
the complex index of refraction \citep{Boh83,Dra84}.

The expansion velocity of the matter can be derived from the
\citet{Vassiliadis93} relationship correlating the expansion
velocity to the pulsational period

\begin{equation}
v_{exp} \left[ km\, s^{-1} \right] = -13.5 + 0.056 P
\label{velocity}
\end{equation}

\noindent with the condition that the velocity is higher than $\rm
3\, km \, s^{-1}$ \citep{Vassiliadis93}. The pulsation period $P$
depends on the pulsation mode of the star.

\noindent
A simple relation between period, mass and radius \citep[eq. (4)
in][]{Vassiliadis93} is obtained supposing that AGB variable stars
pulsate into the fundamental mode:

\begin{equation}
\log P = -2.07 + 1.94 \log \left( \frac{R}{R_{\odot}} \right) -0.9
\left( \frac{M}{M_{\odot}} \right)
\label{period}
\end{equation}

\noindent
where the stellar radius $R$ and mass $M$ are expressed in solar
units.

Another variable that appears in eq. (\ref{tauter}) is the dust-to-gas
ratio $\delta$. We need to relate this ratio to the stellar
parameters. \citet{Hab94} studying the dependence of the gas outflow
velocity $v_{exp}$ at large distances from the star, found that
$v_{exp}$ depends on three parameters, that is the stellar luminosity
$L$, the mass-loss rate $\dot M$, and the dust-to-gas ratio
$\delta$. The observational data indicate that $v_{exp}$ increases at
increasing any one of these parameters as also confirmed by the models
of \citet{Hab94} who also suggested a plausible correlation between
$v_{exp}$ and the dimensions $a$ of the grains. Therefore,
$v_{exp}=v_{exp} (L,\delta,\dot M,a) $ is a function of four
variables. However, $v_{exp}$ is found to depend only weakly on the
mass-loss rate $\dot M$ except when $\dot M$ is small and the same
weak dependence holds good for the grain dimension $a$. See
\citet{Hab94} for all details. Since our aim is to include the effects
of the dust shell around AGB stars on the spectra of SSPs, stars with
a thick shell and high mass-loss rates give the dominant contribution
to IR spectrum of the SSP. Therefore, we adopt here the relation of
\citet{Hab94} for high mass-loss rates, and neglect the dependence of
$v_{exp}$ on $\dot M$ and $a$. The expression for $v_{exp}$ in
presence of high mass-loss rates is

\begin{equation}
  v_{exp} \propto L_{*}^{0.3} \delta^{0.5}
\label{vout}
\end{equation}

\noindent
Following \citet{Bgs98}, who made use of the results by
\citet{Hab94}, we recast $v_{exp}$ as

\begin{equation}
 v_{exp} \simeq 15 \left[ km \, s^{-1} \right]
 \left( \frac{L}{10^{4}L_{\odot}}
 \right)^{0.3} \left( \frac{\delta}{\delta_{AGB}} \right)^{0.5}
\label{voutbis}
\end{equation}

\noindent
and eventually invert it to estimate $\delta$. The factor
$\delta_{AGB}$ depends on the kind of star under consideration.  For
the O-stars we use $\delta_{AGB}=0.01$ as in \citet{Bgs98} and
\citet{Suh99}, whereas for the C-rich stars we adopt the value
$\delta_{AGB}=0.0025$ computed by \citet{Bla98}.

\section{Mass-loss rates from AGB stars}\label{par:massloss}

The rate of mass-loss along the AGB is a key parameter for
the evolution of the stars in this phase because it affects their
lifetime and luminosity.

Despite the great effort devoted to clarify the role of
mass-loss on the evolution of AGB stars, the mechanism of
mass-loss itself is not completely clear. The determinations of
the mass-loss rate from IR and radio data, are affected by
significant uncertainties, due to the poorly known distances,
dust-to-gas ratio, and expansion velocity of the sources.

There are two nearly direct observational evidence for mass-loss:
the continuum IR emission in excess to what we expect from a star
with the typical effective temperature of an AGB, and the
molecular rotational lines detected in emission. The IR emission
observed is characteristic of a dusty shell that absorbs the light
of the star and then emits it in the IR and radio range. The
molecular lines come from the gas surrounding the star: their
widths and time variations prove that the gas is flowing away from
the star.

It is now widely accepted and supported by hydro-dynamical models
that large amplitude pulsations are essential for accelerating the
mass outflow from the stellar surface of AGB stars until the gas
becomes cool enough that heavy elements can condense into dust
grains. The dust grains in turn absorb and scatter the radiation
transferring by collisions energy and momentum from the stellar
radiation field to the gas so that the flow velocity may exceed
the escape velocity \citep{Gil72}. Mass-loss grows with time until
the so-called super-wind regime sets in, which quickly turns the
star into a planetary nebula by stripping away all the envelope
and leaving a bare core that evolves to high temperatures.

\citet{Vassiliadis93}  represented the above situation with
suitable analytical fits in which  $\dot{M}$ exponentially grows
with the luminosity up to the formation of the planetary nebula.
Following the formalism of \citet{Vassiliadis93} we have adopted
the relations below to express the rate of mass-loss prior and
during the super-wind regime. The transition occurs at a period of
about 500 days. Furthermore, once the super-wind regime has
started the stellar envelope gets so rich in dust that the star is
no longer observable in the visible range of the spectrum.

Prior to super-wind, the mass-loss rate is expressed by

\begin{equation}
\log \dot{M}  = -11.4 + 0.0123 P
\label{ratepe}
\end{equation}

\noindent
where $\dot M$ is in $M_{\odot}\, yr^{-1}$ and the period is in
days. During the super-wind regime $\dot{M}$ is assumed to be given by

\begin{equation}
\dot{M} = \frac{1}{cv} \left( \frac{L}{L_{\odot}} \right)
\label{ratepe2}
\end{equation}

\noindent
which describes a wind driven by radiation pressure \citep{Ive94}.

As already pointed out by \citet{Vassiliadis93} long period
variables detected in the optical are found up to periods of $750$
days (top upper part of the  AGB where the stellar mass  is $ \sim
5M_{\odot}$). In order to take this into account,
\citet{Vassiliadis93} slightly changed the expression for the
mass-loss rate of eq. (\ref{ratepe}) according to

\begin{eqnarray}
\log \dot{M}  = -11.4 + 0.0125 \times \left[ P - 100 \left(
\frac{M}{M_{\odot}} - 2.5 \right) \right]
\label{ratepe3}
\end{eqnarray}

\noindent
for stars with masses $M>2.5M_{\odot}$. Once again $\dot M$ is in
$M_{\odot}\, yr^{-1}$, and the period is in days. The equations
(\ref{ratepe}), (\ref{ratepe2}) and (\ref{ratepe3}) completely
describe our mass-loss prescriptions for AGB stars.

\section{Formation and properties of the dusty circumstellar envelopes}\label{par:dustoptical}

{\bf O-rich stars}. O-rich AGB stars of M spectral type show two
typical features at $10 \mu m$ and $18 \mu m$ either in absorption
or in emission depending on the optical depth of the surrounding
envelope. These features are usually attributed to stretching and
bending modes of $Si-O$ bonds and $O-Si-O$ groups and clearly
probe the existence of silicate grains in the shell of matter
around the star.

Among AGB stars, the OH/IR stars are generally thought to represent
the final evolutionary stage of an O-rich object, just before it
evolves quickly into a planetary nebula.  Conversely, Mira variables
of M spectral type are thought to correspond to early or advanced AGB
evolutionary stages of oxygen-rich stars. An oxygen-rich star is
characterized by the ratio $C/O < 1$. Because of the strong triple
bond between O and C in the carbon monoxide, it is believed that all
$C$ will be blocked into $CO$ molecules and no $C$ is available to the
formation of dust grain with other elemental species of low abundance.
In contrast, the fraction of $O$ not engaged in $CO$ reacts with other
elements such as $Mg$ and $Si$ and forms various types of
compounds. Therefore, an O-rich star is characterized by the presence
of $O$ and compounds like $MgO$, silicates and $H_{2}O$. Eventually,
these molecules can bind together to produce grains of silicates.

The newly formed dust grains leave the envelope of the AGB stars and
disperse into the interstellar medium, where they can be strongly
modified by chemical and physical processes, such as collisions,
interaction with energetic photons, accretion and destruction of the
mantle, and so forth \citep[see ][ for a recent review and referencing
of the subject]{Li02}.  Therefore the opacities of dust silicates in
AGB stars are probably different from those derived from laboratory
measurements of terrestrial or meteoritic material.  Furthermore, the
classical opacities of silicates of the diffuse interstellar medium
cannot be used to model the dusty envelopes of AGB stars
\citep{Oss92}.

There are two final points to consider: i.e. the consistency of the
opacities in use with the Kramers-Kronig dispersion relations
\citep[see e.g. ]{Lan60} which imply that the real and imaginary
parts of the complex dielectric function
$\epsilon\left(\omega\right)=\epsilon_{1}+i\epsilon_{2}$ are not
independent, see also \citet{Boh83}, and the consistency of the
theoretical results with the observational ones. In our case, the
spectra obtained from radiative transfer models of  AGB stars
surrounded by dusty shells must agree with the IR observations of
these stars. For more details on this subject see  \citet{Dra84}.

\begin{figure}
\resizebox{\hsize}{!}{\includegraphics{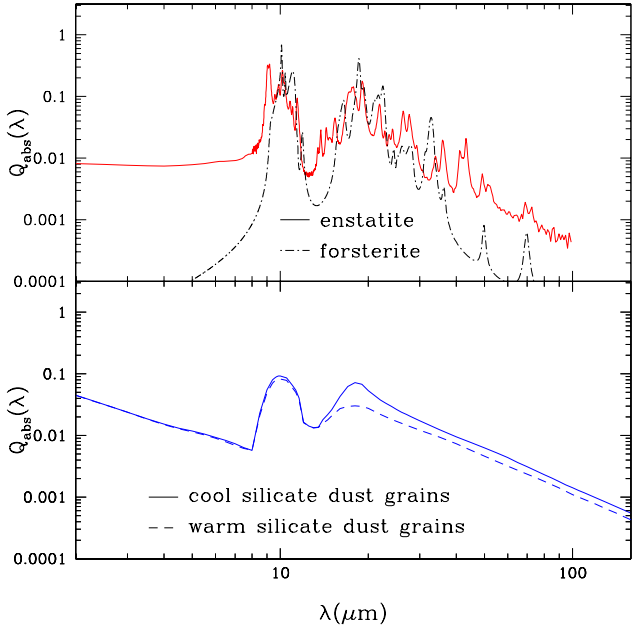}}
\caption{Upper panel: extinction coefficients for two crystalline silicates i.e.
forsterite from Jaeger (2002, private communication) and enstatite
from \citet{Jag98}. Bottom panel: extinction coefficients for cool and
warm silicates taken from \citet{Suh02}.}
\label{Optical}
\end{figure}

As already pointed out by \citet{Suh99}, the dust opacities are often
adjusted in such a way that observations (spectra) are reproduced,
however, without checking for the Kramers-Kronig dispersion relation
at the same time \citep{Vol88,Gri93,SuhJones97}. In some cases
\citep{Oss92} the physical consistency of the input opacity is
secured, but no comparison between observational data and theoretical
predictions is made. A rare exception is \citet{Suh99} who presents
optical constants that satisfy both the Kramers-Kronig relation and
properties of O-rich AGB stars better than in previous studies. In
addition to this, \citet{Suh99} gives two estimates for the silicate
opacities, one for the warm grains (obtained reproducing the emission
feature at $10\mu m $ of OH/IR stars) and one for cold grains
(obtained reproducing the absorption feature at $10\mu m $ always of
OH/IR stars). The variation of the silicate opacity with the
observational constraint to be reproduced can be understood by
considering that AGB stars can be surrounded by thin or thick shells
and that the grain temperature is likely to be cooler in thick shells
than in thin shells. A simple argument supporting this possibility is
that in AGB stars surrounded by a thin shell of matter a large part of
the dust will be at high temperatures, whereas in AGB stars with a
thick circumstellar shell, dust will be at lower temperatures.  Basing
on this, \citet{Suh99}, argues that different kinds of grain
(opacities) are required at varying optical depth of the shell: cold
silicates are suited to thick shells with the $10\mu m$ feature in
absorption, whereas warm silicates are more appropriate to thin shells
with the $10\mu m$ feature in emission.  The largest difference
between the opacities of the two species occurs at $\lambda > 13 \mu
m$ (see the top panel of Fig.~\ref{Optical}) and according to
\citet{Suh99} the transition between the two physical situations
occurs at the optical depth $ \mathcal{T}_{10} = 3$ and $\lambda=10
\mu m$.  The same values are adopted here.

The mid-infrared spectra of oxygen-rich circumstellar envelopes
are dominated by the silicate bands at about $10$ and $19 \mu m$
and so much less attention has been paid to the other IR spectral
regions until the advent of IRAS and ISO in particular.  However,
in many IRAS LRS spectra of Mira variables \citep{Slo96} an
emission band at about $13 \mu m$ has been detected that could be
due to aluminum oxide \citep{Beg97}, but its origin is still a
matter of vivid debate \citep{Pos99}. Other weak silicate features
around 10 and 18 $\mu m$ have been detected by the ISO-SWS
observations which can be attributed to the presence of oxide
particles. In many AGB stars with high mass-loss rates, ISO high
resolution observations revealed the presence of prominent bands
of crystalline silicates, for instance enstatite ($MgSiO_{3}$) and
forsterite ($Mg_{2}SiO_{4}$) \citep{Wat96,Wat99}.

All these features spurred the interest toward the composition of
circumstellar dusty shells and grain formation theories
\citep{Gai99}. \citet{Suh02} presented a dust model for the envelopes of O-rich
stars in which not only amorphous silicates but also crystalline
silicates are included. The adopted opacity functions are those of
\citet{Jag98} (the bottom panel of Fig.~\ref{Optical}). The key
parameter controlling an envelope composed by amorphous silicates
and crystalline silicates is the so-called {\it crystallinity}
parameter $\alpha$, that is the ratio between total mass of
crystalline silicates to the total mass of silicates.
\citet{Suh02}, examining various levels of crystallinity, found
that for the same value of $\alpha$, crystalline silicate features
are stronger for optical depths smaller than 5 and weaker for
optical depths larger than 10 (always referred to the optical
depth at $10 \mu m$). A high crystallinity is required to produce
strong features in models with high optical depth.

The intensity of the crystalline silicate features is likely to
correlate with the mass-loss rate from AGB stars.  As a matter of fact
these features are stronger in OH/IR stars with high mass-loss rates
\citep{Sylvester99}. Because the observations show  that the same
features are not present in stars with low mass-loss rates, one may
perhaps infer that these stars are also deficient of crystalline
silicates in their envelopes \citep{Suh02}.  The problem, however,
needs deeper analysis because \citet{Suh02} and \citet{Kemper01} find
contrasting results. \citet{Kemper01} even including the presence of
crystalline silicates in the dusty envelopes, fail indeed to reproduce
the features in question.

Following \citet{Suh02}, we adopt here the crystallinity parameter
$\alpha = 0.1$ for stars with low mass-loss rates and moderately
optically thick shells of matter ($\mathcal{T}_{10} < 15$), whereas
for OH/IR stars with high mass-loss rates and optically thick shells
($\mathcal{T}_{10} > 15$) we prefer the value $\alpha = 0.2$. It is
worth recalling that $\alpha = 0.2$ is fully sufficient to reproduce
the prominent crystalline features shown by OH/IR spectra. Finally, in
all the models the relative contents of enstatite $\left(
MgSiO_{3}\right)$ and forsterite $\left( Mg_{2}SiO_{4}\right)$ are the
same as in \citet{Suh02}, i.e. we adopt the same relative contribution
for the two components.

{\bf C-rich stars}. C-stars are the evolutionary descendants of Mira
variables. Through continuous dredge-up of carbon into the envelope
during the thermally pulsing AGB phase, these stars eventually reach a
carbon abundance in the outer layers larger than that of oxygen ($C/O
> 1$). When this occurs the formation of O-rich dust ceases and it is
replaced by that with C-rich compounds: the C-star phase
begins. Therein after, the continuous formation of C-rich dust makes
the envelopes of these stars more and more optically thick. By loosing
mass at very high rates, they get enshrouded by thick envelopes that
absorb and scatter the UV-optical radiation into the IR and radio
range.

\begin{figure}
\resizebox{\hsize}{!}{\includegraphics{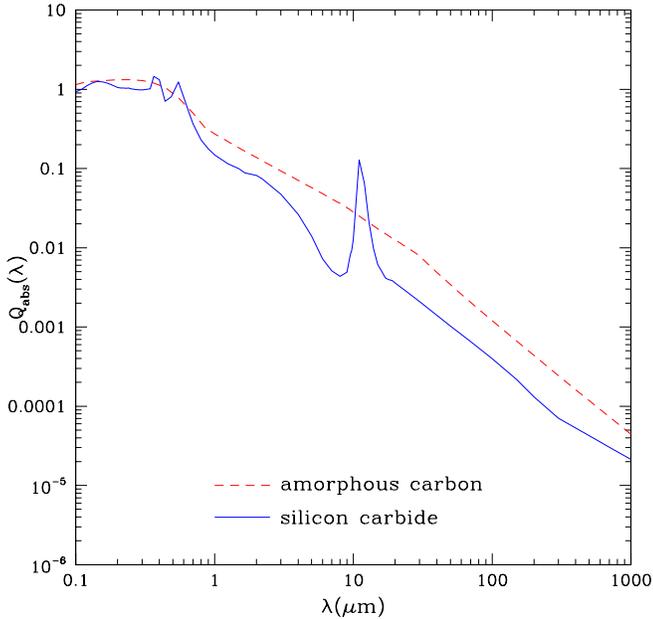}}
\caption{Absorption coefficients for amorphous carbon (dashed
line) from \citet{Suh00} and silicon carbide (solid line) from
\citet{Peg88}.}
\label{Optical2}
\end{figure}

Almost all these stars show an emission feature at $11.3 \mu m$,
due to silicon carbide ($SiC$), whose presence was predicted by
\citet{Gil69} and observationally confirmed by  \citet{Hac72}.
Many infrared observations have brought into evidence and
confirmed the presence of several types of dust grains in C-rich
AGB stars: amorphous carbon ($AMC$), silicon carbide, and
magnesium sulphide ($MgS$) to mention the three dominant types.
\citet{Suh99,Suh00} has derived new opacities for the $AMC$, that
are consistent with Kramers-Kronig dispersion relations, and has
also reproduced the observational data by means of suitable models
of the dusty shells around AGB stars (see Fig.~\ref{Optical2}).
The models are based on the new opacities and a complete treatment
of the radiative transfer, thus much improving upon previous
studies \citep{Bla98,Gro98}. The \citet{Suh99,Suh00} models are
characterized by two components, $SiC$ and $AMC$ (for typical
models of carbon stars, see for example \citet{Lor01} and
references therein). The $30 \mu m$ $MgS$ feature is observed in a
wide range of sources, going from low mass-loss carbon stars to
planetary nebulae \citep{Hon02}. Because it is observed in C stars
of large luminosity it is expected to significantly contribute to
the integrated spectrum in the $30 \mu m$ range. However, as the
optical constants of $MgS$ are not measured for a sufficiently
wide range of wavelengths \citep{Suh00,Hon02} the presence of
$MgS$ is neglected here.

The grains of silicon carbide can be separated in two groups,
hexagonal or rhombohedral $\alpha SiC$ and cubic $\beta SiC$. Many
authors \citep{Gro95,Spe97,Bla98} argue that in order to fit the IR
spectra of C-stars, $\alpha SiC$ is required sometimes together with
$\beta SiC$.  However, all studies of meteoritic $SiC$ grains have
revealed the presence of only $\beta SiC$. Since it is
thermodynamically unlikely that $\alpha SiC$ transforms itself into
$\beta SiC$ \citep{Ber97}, there seems to be a clear discrepancy
between the $\alpha SiC$ suggested by the AGB spectra and the $\beta
SiC$ indicated by the meteoritic data. As pointed out by \citet{Spe99}
the discrepancy could be due to inadequate KBr corrections
\footnote{\citet{Dor78} have studied the effects of the a dispersive
medium made of potassium bromide (KBr) on the spectra of small
quantities of silicates with the grain size.  They find a shift of the
spectral features emitted by the grains using either a KBr-matrix or a
KBr-pellet in which the small grains are embedded. They explain the
shift as due to an effect caused by the KBr-matrix that needs to be
corrected before using the spectra obtained with the KBr-pellet. See
\citet{Spe99} for more details about the problem of the KBr
correction.} to the laboratory spectra so that a plausible way out
can be found.

$AMC$ and $SiC$ affect the spectrum in a different fashion: while the
effects of $AMC$ propagate over the whole spectrum, those of $SiC$ are
limited to the $11 \mu m$ feature as observationally
indicated. \citet{Lor94} and \citet{Gro95} suggest that the ratio
$SiC$ to $AMC$ decreases at increasing optical depth of the dusty
envelope. This is also confirmed by the models of \citet{Suh99}.
Therefore, following \citet{Suh99} prescription, we assume here that
the chemical composition of the dust changes with the optical depth of
circumstellar shell. For optically thin dust shells ($\mathcal{T}_{10}
\leq 0.15$), the strong $11 \mu m$ feature requires about $20 \%$ of
$SiC$ dust grains to fit the observational data for AGB; for dust
shells with intermediate optical depth ($0.15 \leq \mathcal{T}_{10}
\leq 0.8$) about $10 \%$ $SiC$ dust grains are needed, whereas for
shells with larger optical depths, in which the $11 \mu m$ feature is
either much weaker or missing at all, no $SiC$ is required. Following
\citet{Suh99}, who used the optical constants of $\alpha SiC$ by
\citet{Peg88} to calculate the opacity for $SiC$ and reproduce the
spectra of C-stars similarly, we adopt here the same data (see
Fig.~\ref{Optical2}).

\section{Parameters of the shell model}\label{par:parameters}

An important parameter of the dusty shell models is the distribution
of the grain dimensions. Nowadays many recipes have been proposed to
reproduce the properties of the dust grains: they go from the
classical power-law \citep{Mat77} to a log-normal distribution of
\citet{Wei01}. In this study we adopt the simple distribution
expressed by the Dirac delta function

\begin{displaymath}
n \left( a \right) = \delta \left( a - a_{0} \right)
\end{displaymath}

\noindent
with $a_{0} = 0.1 \mu m$, where $a$ is the dimension of the
grains. Despite its simplicity, this type of distribution is fully
adequate to our purposes and it has already been widely used to model
the dusty shells of AGB stars
\citep{Has95,Ive95,VdV95,SuhJones97,Suh97,Suh99,vLo99,Suh00,Suh02}.
Support to this kind of approximation is by \citet{Bag95} who find
that a delta-function fits the spectrum of IRC+10216 much better
than a power law.

Since different types of grains are present in the dust mixture of the
circumstellar shell, this could imply different values of the grain
temperature at the same distance from the central source. As a
consequence of it, the condensation temperature at inner boundary of
the shell could vary with the type of grain. In the case of shells of
the C-rich stars, for which we use a mixture of $AMC$ and $SiC$,
$T_{d}$ is assumed equal to $1000 °K$ for both types of grain: in
reality $AMC$ and $SiC$ should have their own temperature and
temperature profile across the shell. This of course would add
additional parameters to the problem, i.e. two condensation
temperatures and two dust-to-gas ratios, thus rendering the whole
problem much more complicated and beyond the technical capability of
the code DUSTY we are using \citep*{Nen99}. The code indeed treats a
mixture of grains using a pseudo-grain whose optical properties are
the average of those of individual species and of course deals with a
single value of $T_{d}$. Fortunately, in the case of C-rich stars, the
abundance of $SiC$ is very low so that the approximation to a single
value of $T_d$ is physically acceptable.  By the same token, we may
also assume the same value of $T_{d}$=1000 $°K$ for this type of
shell. As first noticed by \citet{Suh99}, if in an expanding stellar
envelope dust grains condense in amorphous form and then get a
crystalline structure on a short time scale, it is likely that two
different types of grain have the same temperature distribution.

Another important parameter is the density profile of the grains
across the shell. Many different density laws have been suggested:
the simple power-law $\rho \propto r^{-x}$ is a popular choice
even if the exponent $x$ varies from author to author. The law
$\rho \propto r^{-2}$ is often adopted. Equally good alternatives
are $\rho \propto r^{-1}$ or $\rho \propto r^{-3}$. The latter
seems more suited to describe the case of O-stars with thick
envelopes. Furthermore, models with radiation pressure or models
with pulsations and shocks predict $\rho \propto r^{-2}$. Finally,
a region of finite thickness with enhanced density is added to the
distribution $\rho \propto r^{-2}$ to somehow describe stars that
are in the super-wind phase \citep{SuhJones97,Suh97}. In this
study we  simply adopt  the power-law $\rho \propto r^{-2}$ as in
the series of papers by \citet{Suh99,Suh00,Suh02} from which we
also take the opacities of the dust grains.

However, using different power-laws for the distribution of the matter
in the shell implies that the same overall optical depth is reached at
different radii. In order to evaluate the uncertainty on the optical
depth $\mathcal{T}_\lambda$ caused by using different density profiles
across the shell.  We start from eq. (\ref{tau}), insert the relation
$\rho_d(r) = \delta \rho(r)$ where $\delta$ is a constant, suppose for
the sake of simplicity that the product $k_\lambda(r)\delta$ does not
depend on $r$, and finally insert the generic radial dependence for
the density $\rho_{0} \left( r / r_{0} \right)^{-n}$ (where $\rho_{0}$
and $r_{0}$ are suitable scale factors).  The optical depth down to a
generic radius $r$ is

\begin{eqnarray}
\mathcal{T}_{\lambda}\left(r \right) &=& \langle k_\lambda(r)\delta
\rangle
           \int_{r_{in}}^{r} \rho(r)  dr \nonumber \\
&& = \langle k_\lambda(r)\delta \rangle  \rho_0
           r_{0}^{n} \int_{r_{in}}^{r} r^{-n}  dr
\end{eqnarray}

\noindent
At any given radius $r$ the variation caused by using different
power-laws is

\begin{equation}
 \Delta \left(r_{0}^{n} \int_{r_{in}}^{r} r^{-n} \, dr\right)
\end{equation}

\noindent
Let us indicate with $\mathcal{I}_1$, $\mathcal{I}_2$ and
$\mathcal{I}_3$ the three integrals for three typical power-law
exponents $n=1, 2$ and $3$, respectively, pose $r=\alpha r_{in}$, and
finally assume the scale length $r_{0}$ to be the inner radius of the
shell $r_{in}$.  We immediately get

\begin{displaymath}
\mathcal{I}_1 = r_{in}ln \alpha \quad \mathcal{I}_2 \simeq
r_{in}\left(1-\frac{1}{\alpha}\right) \quad \mathcal{I}_3 \simeq
0.5r_{in}\left(1-\frac{1}{\alpha^{2}}\right)
\end{displaymath}

\noindent
The factors multiplying $r_{in}$ give an idea of the differences due
to the power-law exponent $n$. The net consequence of different mass
density profiles is that the more matter is stored at large radii and
hence lower temperatures the flatter is the spectrum and vice-versa
\citep[see][ for more details]{Ive97}.

\section{Uncertainties of the shell model }\label{uncerta}

In this section, we try to discuss the uncertainties arising from
the two main components of the adopted model for the dust shell.

{\it Optical depth.} Differentiating eq. (\ref{tauter}) we get

\begin{equation}
{\Delta\mathcal{T}_{\lambda} \over \mathcal{T}_{\lambda} } =
{\Delta\delta \over \delta} +
{\Delta \dot M\over \dot M } +
{\Delta k_\lambda \over k_\lambda} +
{\Delta v_{exp} \over v_{exp}} +
{\Delta r_{in} \over r_{in}}
\label{diff_tau}
\end{equation}

\noindent
In the following we try to estimate the various terms composing this
relation. If we look at eq. (\ref{voutbis}), this one well represents
the link between luminosity of the star $L_{*}$, outflow velocity
$v_{exp}$ and dust-to-gas ratio $\delta$, because \citep{Hab94} if
$L_{*}$ or $\delta$ increase, also $v_{exp}$ increases, but the
exponents that rule the relation of proportionality are uncertain. We
adopt the dependence proposed by \citet{Hab94}, but for example
\citet{vLo00} proposed a slightly different choice for the $L$
dependence. If we write eq.  (\ref{voutbis}) as $v_{exp} \propto
L_{*}^{\alpha} \delta^{\beta}$ we get

\begin{equation}
{\Delta\delta \over \delta} = \frac{1}{\beta}{\Delta v_{exp} \over
v_{exp}} + \frac{\alpha}{\beta} {\Delta L \over L}
\end{equation}

\noindent
where $\alpha$ and $\beta$ are $2$ and $0.6$ in our model. The errors
$\Delta v_{exp}$ and $\Delta L$ are small, because equations
(\ref{velocity}) and (\ref{period}) are sufficiently good and so, if
we consider a large spread of $\alpha$ and $\beta$ and typical values
for the expansion velocities and luminosities of AGB stars, we get
${\Delta\delta /\delta} \simeq 1$ in the worse cases.

The uncertainty on the rate of mass-loss can be estimated from eq.
(\ref{ratepe}) prior to super wind and eq. (\ref{ratepe2}) during the
super-wind. Differentiating eq. (\ref{ratepe}) we obtain

\begin{equation}
{\Delta \dot M \over \dot M } = {1\over log_{10}e} 0.0123 \Delta P
  \simeq 0.034 \Delta P
\end{equation}

\noindent
which for the error on the period coming from (\ref{period}) yields
that ${\Delta \dot M /\dot M }$ is negligible. The uncertainty on
eq. (\ref{ratepe2}) is simply

\begin{equation}
{\Delta \dot M \over \dot M }= {\Delta L \over L } +
                               {\Delta v_{exp}  \over v_{exp} }
\end{equation}

\noindent
which is also negligible.

\noindent
Estimating the term ${\Delta k_\lambda / k_\lambda}$ is a cumbersome
affair because $k_\lambda$ depends on the mass abundances $\chi_i$ of
the grains, which in turn are functions of the optical depth
$\mathcal{T}_\lambda$ and of the cross sections $\sigma_{g_{i}}$s of the
dust-radiation interactions, which depend in its turn on the grain
dimensions $a$ and on the extinction coefficients $Q_{ext}(i)$. No
simple way of checking the uncertainty can be found, but referring to
the original sources of data for extinction properties.

As already mentioned, the uncertainty on the expansion velocity
$v_{exp}$ is simply estimated from eq. (\ref{velocity}), which yield a
negligible ${\Delta v_{exp} / v_{exp}}$.

Finally, the uncertainty on the inner radius of the shell can be
derived from differentiating eq. (\ref{inner_shell})

\begin{equation}
{\Delta r_{in} \over r_{in}} =
 2 {\Delta T_d \over T_d}
 + 2 {\Delta L \over L}
\end{equation}

\noindent
For $T_d$ we have many different values in literature: varying it in
the realistic range from $1000$ to $1500 K$ and taking the typical
luminosity for AGB stars we estimate ${\Delta r_{in} / r_{in}} \simeq
1$. We can finally conclude that the terms ${\Delta r_{in} / r_{in}}$
and ${\Delta\delta / \delta}$ are the main source of error in the
calculation of the optical depth $\mathcal{T}_\lambda$. For individual
stars the error may be as large as ${\Delta\mathcal{T}_{\lambda} /
\mathcal{T}_{\lambda} } \simeq 2$ which is acceptable for the purposes
of our study. It will be smaller as soon as better determinations of
the relationship $v_{exp}$($L_{*}$, $\delta$) will be available.

{\it Chemistry.} Although the list compounds we have considered in
modelling the chemical composition of the dust shells is certainly
incomplete, an effort has been made to include as many molecules
as possible. Many other compounds are of course possible and
should be included in the list, but we need deeper studies able to
discriminate for which AGB stars these compounds have to be
included. However, compared to similar studies on the same topic,
e.g. \citet{Bgs98}, our treatment is a significant step forward.
To conclude we are confident that our description of the shell
composition is adequate to the present aims. No doubt that this
aspect can be improved.

{\it Final remarks.} Even if the uncertainty is as large as
${\Delta\mathcal{T}_{\lambda} / \mathcal{T}_{\lambda} } \simeq 2$,
the situation is, however, not as bad as it may seem, because the
largest effect of the dust shell on re-processing the radiation
coming from the central star underneath is for the optically thick
case, in which the exact value of the optical depth is less of a
problem.

\section{Theoretical SSPs: basic ingredients}\label{theo_SSP}

The monochromatic flux of an SSP of age $t$ and metallicity $Z$
at the wavelength $\lambda$ is defined as

\begin{equation}
ssp_{\lambda }\left( t,Z \right) =\int\nolimits_{M_{L}}^{M_{U}(t)} f_{\lambda }\left(
M,t,Z \right) \Phi (M) dM
\label{def_flux_ssp}
\end{equation}

\noindent where $f_{\lambda }\left( M,t,Z\right) $ is the
monochromatic flux emitted by a star of mass $M$, age $t$, and
metallicity $Z $; $\Phi\left(M\right)$ is the initial mass
function (IMF); $M_{L}$ is the mass of the lowest mass star in the
SSP whereas $M_U(t)$ is mass of the highest mass star still alive
in the SSP of age $t$. For the IMF we adopt the \citet{Salpeter55}
law expressed as

\begin{displaymath}
{dN \over dM }= \Phi(M) = \mathcal{A} M^{-x}
\end{displaymath}

\noindent
where $x$=2.35 and $\mathcal{A}$ is a normalization constant to be
fixed by a suitable condition (SSPs for other choices of the IMF can
be easily calculated).

Because the flux $ssp_{\lambda }\left( t,Z \right)$ is calculated
by integrating equation (\ref{def_flux_ssp}) along an isochrone,
one has to know the luminosity, $T_{eff}$ and gravity of the stars
of mass $M$, age $t$ and metallicity $Z$ lying on the isochrone.
For any star of mass $M$, age $t$ and metallicity $Z$, the
relationship between luminosity, $T_{eff}$, gravity with the age
$t$ is derived from a library of stellar models, the flux
$f_{\lambda }\left(M,t,Z \right)$ from a library of stellar
spectra. Finally one has to adopt a prescription for the amount of
mass lost by a star of mass M in the course of its evolution: this
could occur all over the evolutionary history such as in case of
massive stars, or at the end of the RGB passing from the tip of
the RGB to the HB or clump as appropriate, and during the AGB
phase as in the case of low and intermediate mass stars.  All
details of the isochrone construction technique can be found in
\citet{Ber94}.

\subsection{Libraries of stellar models  and stellar spectra}\label{par:models_spectra}

In our study we adopt the isochrones by \citet{Tantalo98a}
(anticipated in the data base for galaxy evolution models by
\citet*{Leitherer96}).  The underlying stellar models are those of the
Padova Library. They are shortly referred to as the \citet{Ber94}
stellar models (see references therein for more details). These
stellar models are calculated with convective overshooting and are
amply described by \citet{Ber94} so that no detail is given here.  In
the present study only three chemical compositions are considered,
namely [Y=0.240, Z=0.004], [Y=0.250, Z=0.008], [Y=0.280, Z=0.02], for
which the transition luminosities for AGB stars passing from the
O-rich to the C-rich regime calculated by \citet{Mar99} are available
(see below).

In the isochrones, all evolutionary phases from ZAMS to the end of
the TP-AGB or C-ignition are included, as appropriate to the mass
of the last living star of age $t$. The rate of mass-loss during
the RGB stages of low mass stars is from \citet{Rei75} with $\eta
= 0.45$. During the AGB phase the rate of mass-loss is according
to \citet{Vas93}, as already described in
Sect.~\ref{par:massloss}. The initial masses of the stellar models
go from 0.15 to 100 $M_{\odot}$ and the ages of the isochrones go
from 3 Myr to 20 Gyr.

The library of stellar spectra is from \citet{Lejeune98}, which stands
on the Kurucz (1995) release of theoretical spectra, however with
several important implementations. For $T_{eff}<3500$ $^{\circ }K$ the
spectra of dwarf stars by \citet{All95} are included and for giant
stars the spectra by \citet{Flu94} and \citet{Bes89,Bes91} are
considered. Following \citet{Bre94}, for $T_{eff}> 50000$ $^{\circ
}K$, the library has been extended using black body spectra.

For ages younger than about 100 Myr the SSPs we are going to present
will be equal to the old ones of \citet{Tantalo98a}, because no AGB
stars can be formed, whereas for older ages the effect of the
circumstellar envelopes in processing the radiation coming from the
central star is taken into account.

\subsection{Transition luminosities of the AGB phases}\label{par:transition}

Owing to the different composition of the ejecta much affecting the
chemistry and physics of the circumstellar shell, the transition from
the oxygen-rich M-type stars ($C/O <1$) to the carbon-rich C-type
objects ($C/O >1$) deserves particular care.

In view of the discussion below, it is useful to summarize the
evolution of AGB stars eventually becoming C-stars. The situation is
illustrated in Figs.~\ref{AGBZ004OC}, \ref{AGBZ008OC} and
\ref{AGBz02OC}  for three different values of the metallicity,
namely $Z$=0.004, $Z$=0.008 and $Z$=0.02 respectively. Each plot shows
the bolometric magnitude of TP-AGB stars as a function of their
initial mass $M_{i}$ from the beginning to the end of the AGB phase
when the outer envelope is completely removed by mass-loss. The thick
dashed and solid line bind the luminosity attained by the stellar
models of the Padova Library \citep{Ber94} used by \citet{Tantalo98a}
assuming the prescription for mass-loss of \citet{Vas93}. The thin
dashed and solid lines show the same according to the more recent
models of TP-AGB stars by \citet{Mar99}. The shaded areas correspond
to the luminosity interval in which, according to \citet{Mar99},
C-stars develop. Examining the three diagrams and the models by
\citet{Mar99} in detail we note that:

\begin{figure}
\resizebox{\hsize}{!}{\includegraphics{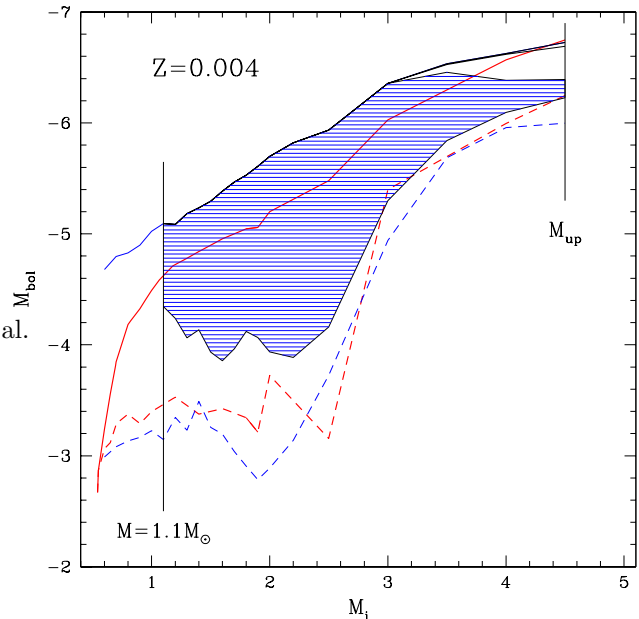}}
\caption{The range of bolometric magnitudes spanned by AGB stars
with $Z$=0.004 and the transition bolometric magnitudes from M- to
C-star as function of the initial mass. The dashed lines show the
start of the AGB phase. The solid lines show the end of it. The shaded
area is the luminosity interval in which C-stars are formed. The thick
lines (dashed and solid) are the models by \citet{Tantalo98a}
calculated with the \citet{Vas93} prescription for mass-loss during
the AGB phase. The thin lines (dashed and solid) show the same but for
the models by \citet{Mar99}. The region of C-stars is according to the
\citet{Mar99} models. The two vertical lines show the minimum mass for
the formation of C-stars and the maximum mass $M_{up}$for the
occurrence of the AGB phase.}
\label{AGBZ004OC}
\end{figure}

\begin{figure}
\resizebox{\hsize}{!}{\includegraphics{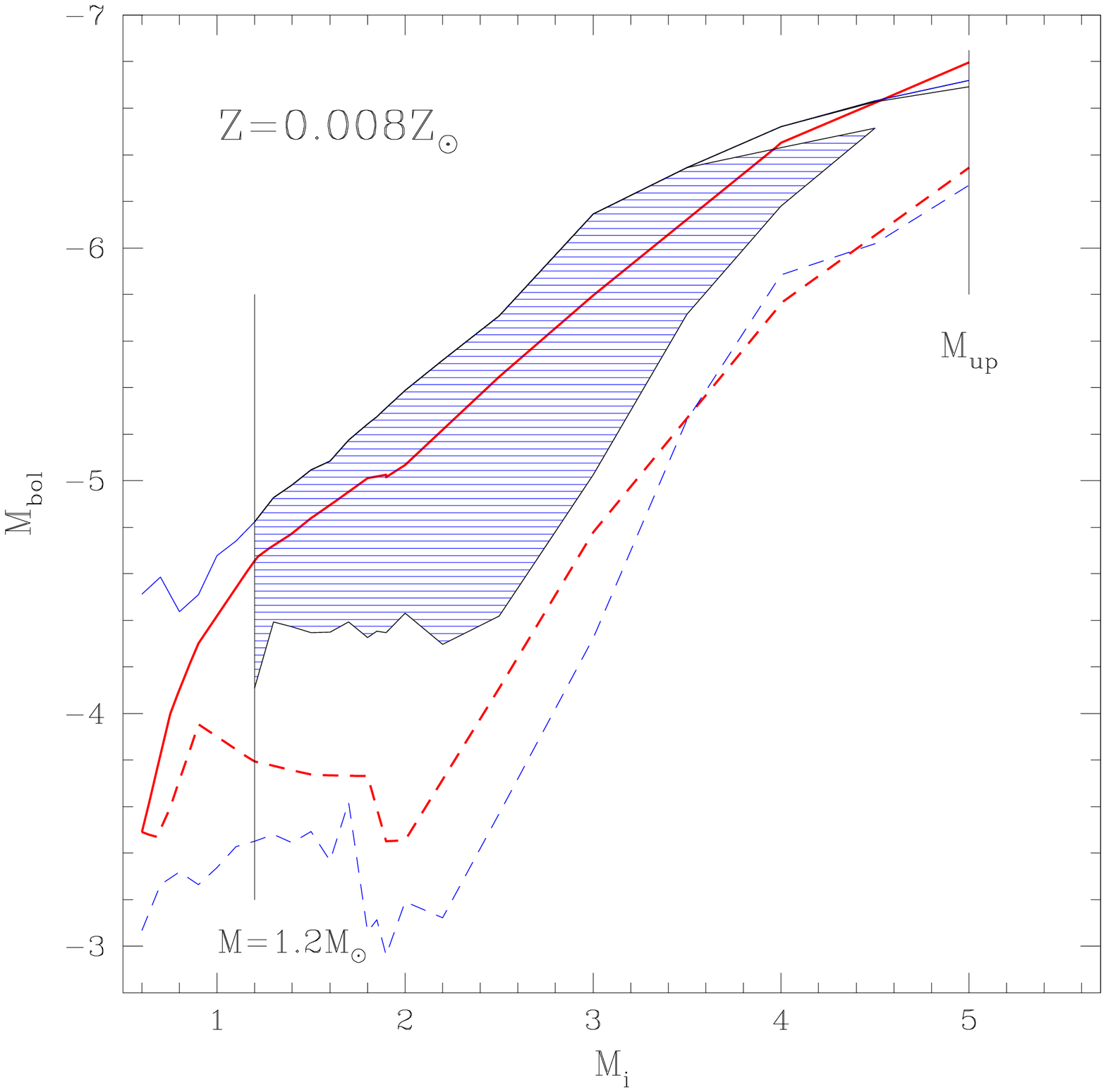}}
\caption{The same as in  Fig.~\ref{AGBZ004OC} but for the metallicity $Z=0.008$.}
\label{AGBZ008OC}
\end{figure}

\begin{figure}
\resizebox{\hsize}{!}{\includegraphics{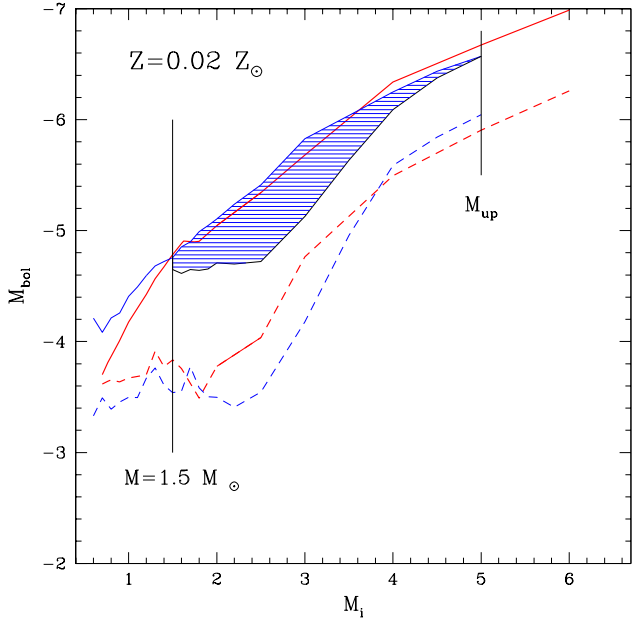}}
\caption{The same as in  Fig.~\ref{AGBZ004OC} but for the metallicity $Z$=0.02.}
\label{AGBz02OC}
\end{figure}

\noindent
(i) Over the mass interval in which the AGB phase develops, the
formation of C-stars does not occur for initial masses lower than a
certain limit indicated by the vertical line (the limit mass increases
at increasing metallicity). This is simply caused by the very early
loss of the envelope before any change in the chemical composition of
the outer layers may take place. Therefore, in old stars only O-rich
shells of dust with $C/O < 1$ are possible. At larger initial masses,
C-stars and C-rich dusty shells do occur over a wide range, which may
extend up the so-called $M_{up}$, i.e the maximum value for the
occurrence of the AGB phase, or slightly below this. In the range of
the massive AGB, the occurrence of envelope burning may indeed delay
the formation of a C-star, and in some cases may turn a C-star into a
late O-rich star. More precisely, in this mass range a C-star is
comprised between two stages, in which an AGB star appears as
O-rich. The late O-rich phase is favored at low metallicities and
disappears for $Z\gtrsim Z_{\odot}$.

\noindent
(ii) As a consequence of the above trend, for the majority of C-stars
(say up to $M_{i} \lesssim 3.5M_{\odot}$) the maximum luminosity just
coincides with the end of the AGB phase, whereas for initial masses
larger than $M_{i} \gtrsim 3.5M_{\odot}$ the maximum luminosity of
C-stars may be fainter than the maximum luminosity of the AGB phase.

Given these premises, we need to incorporate the results by
\citet{Mar99}, which provide the transition luminosity into the
isochrones (SSPs) by \citet{Tantalo98a}. The comparison of the
initial, transition, and maximum luminosities in the two sets of
models reveals that:

\noindent
(a) For $Z=0.004$ the luminosity at which the TP-AGB phase begins is
nearly the same for both \citet{Tantalo98a} and \citet{Mar99}, whereas
the termination luminosity is lower in \citet{Tantalo98a}. Finally
C-stars are formed for initial masses larger than about
$1.1M_{\odot}$. Owing to the lower termination luminosity, the
proportion of C-stars with respect to that of the O-rich ones
predicted by the \citet{Tantalo98a} will be somewhat smaller compared
to the value expected from \citet{Mar99}.

\noindent
(b) Similar considerations apply to case with $Z=0.008$, the only
difference being that no C-stars can be formed for initial masses
lower than about $1.2 M_{\odot}$. We note that the range of bolometric
magnitude predicted by the models of \citet{Tantalo98a} is narrower
than that by \citet{Mar99}, with lower maximum and higher initial
luminosities for the AGB phase. This reduces the allowed range for
both M- and C-stars. Nevertheless, to a first approximation the
percentage of C-stars with respect to that of O-rich stars remains
similar to that predicted by \citet{Mar99}.

\noindent
(c) For $Z$=$Z_{\odot}$, the initial and maximum luminosity of the
TP-AGB phase in \citet{Tantalo98a} and \citet{Mar99} are nearly the
same so that there would be no difference using either
\citet{Tantalo98a} or \citet{Mar99} models. Now the formation of C-stars
occurs for initial masses larger than 1.5$M_{\odot}$.

\noindent
(d) Finally, for the sake of simplicity the possibility of a late
O-rich phase is taken into account only for $Z$=0.004 and $Z$=0.008
and is ignored at higher metallicities owing to the very small
difference between the maximum luminosity of the AGB phase and of the
C-star regime.  See for instance the case of $Z$=0.02 in
Fig.~\ref{AGBz02OC}.

Basing on the above considerations the use of the \citet{Tantalo98a}
isochrones is fully adequate to our aim of studying the effect of the
dusty envelope on processing the radiation coming from the central
object. Therefore we will adopt the \citet{Tantalo98a} models as
source of our SSPs and use the models by \citet{Mar99} only to fix the
luminosity at which the transition from O-rich to C-stars occurs. Work
is in progress to replace the SSPs by \citet{Tantalo98a} models with
new ones fully incorporating the TP-AGB stars as modeled by
\citet{Mar99}.

\subsection{A remark on the libraries of stellar spectra and the evolutionary
models of AGB stars}\label{spectra_agb}

The general accuracy of the final result depends on the adopted
libraries of stellar spectra, in particular the correlation between
the theoretical parameters $T_{eff}$, gravity and chemical abundance
and the associated SED, and important details of the evolutionary
models for AGB stars, in particular the dredge-up episodes altering
the surface chemical abundances, the efficiency of mass-loss
terminating the AGB evolution, the path of AGB stars in the HR-Diagram
eventually determining the $T_{eff}$ and luminosity of the AGB stars
and the associated SEDs in turn, and finally the accuracy of the
isochrone construction and of the stellar models underneath. To
summarize and review in detail all the sources of the various items
above would lead to too a lengthy discussion which goes perhaps beyond
the scope of this study. Suffice it to mention, that the libraries of
stellar spectra, isochrones, SSPs in usage here have been successfully
tested against, somehow tailored to match, and applied to study the
broad band UBVRI photometric data of star clusters (stellar
populations in general), e.g. \citet{Gir98}, \citet{Gir99},
\citet{Carraro99}, \citet{Tantalo98b}. As far as the evolutionary models of AGB stars
are concerned, the same considerations hold good, because they have
been calibrated on the observational data, e.g. the luminosity
function of C-stars in the Magellanic Cloud \citep{Mar99}. To conclude
the libraries of stellar spectra and evolutionary models of AGB stars
we have adopted are fully adequate to our purposes. It is, however,
long known that the IR colors of the classical models of AGB stars
disagree with the observational data. In addition to improving the
physics of the models, for instance by including the effect of C-rich
composition on the opacity of the outermost layers, see \citet{Mar02}
which would yield AGB models of much cooler $T_{eff}$ during the
C-rich phase (these models are not yet incorporated in this study),
the dust shells are expected to strongly affect the IR colors and to
bring them to better agree with the observations. This indeed is the
aim of this study.

\section{Infrared spectra of SSPs}\label{par:SSPs}

\begin{figure}
\resizebox{\hsize}{!}{\includegraphics{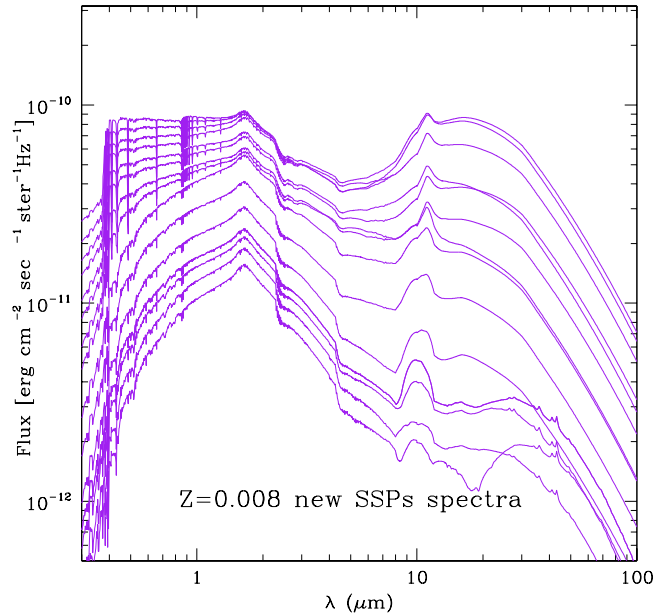}}
\caption{Integrated SEDs  $F_{\nu}$ vs. $\lambda$ for the  SSPs
with $Z=0.008$, ages from 0.25 to 10 Gyr and the inclusion of dusty
circumstellar envelopes in AGB stars. From the bottom to the top the
displayed ages are : 10, 7.5, 5, 4, 3, 2, 1.5, 1, 0.8, 0.6, 0.5, 0.4,
0.35, 0.3, and 0.25 Gyr.}
\label{FluxSSPZ008}
\end{figure}

\begin{figure}
\resizebox{\hsize}{!}{\includegraphics{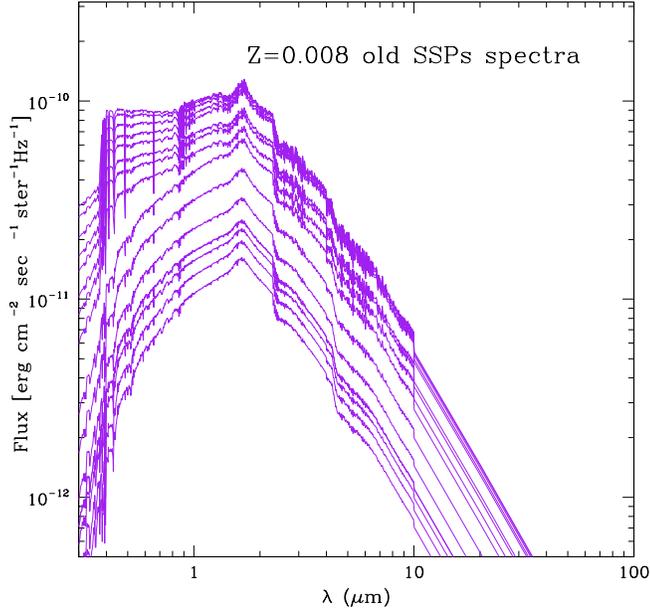}}
\caption{ The same as in Fig.~\ref{FluxSSPZ008} but for the
classical SSPs of \citet{Tantalo98a}.}
\label{FluxOLDSSPZ008}
\end{figure}

\begin{figure}
\resizebox{\hsize}{!}{\includegraphics{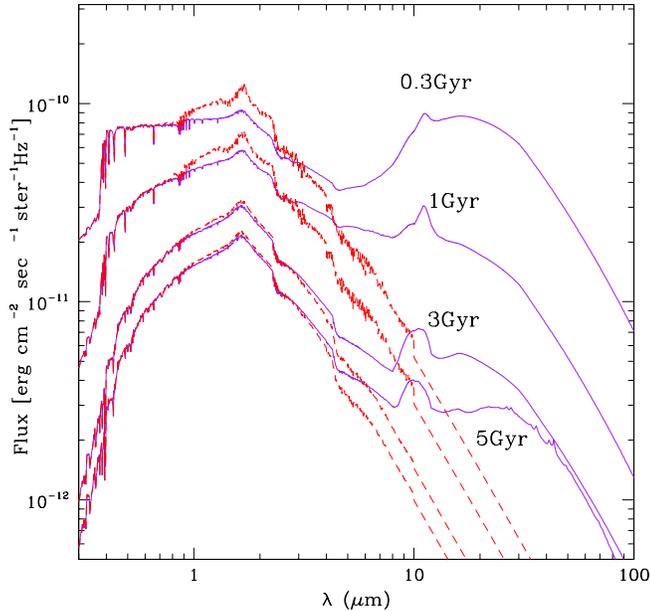}}
\caption{Detailed comparison of SEDs  $F_{\lambda}$ vs. $\lambda$
for the old (dashed lines) and new SSPs (continuous lines) with
$Z=0.008$. Only four ages are considered as indicated.}
\label{FlSSPONZ008}
\end{figure}

In Fig.~\ref{FluxSSPZ008} we plot the SED of dusty SSPs for different
values of the age limited to the case with metallicity $Z$=0.008 and
in Fig.~\ref{FluxOLDSSPZ008} we show the same but for the classical
SSPs in which the effect of dust is neglected.  The displayed SEDs are
for ages in the range 0.25 to 10 Gyr.  A better view of the difference
brought by dust is shown in Fig.~\ref{FlSSPONZ008}, in which the new
and old SEDs are compared for a few selected ages.  The differences
are remarkable. First of all, in the old SSPs the spectra do not
extend into the medium and far IR (MIR and FIR, respectively), but
sharply decline for wavelengths longer than about $3-4 \, \mu m$.  In
contrast, the spectra of the new SSPs extend toward long wavelengths
and the flux is considerable also in the MIR and FIR.

The differences start at about $1\mu m$ and in the IR range up to
$3-4\, \mu m$ the flux of dusty SSPs is lower than the old one: this
is a consequence of the fact that dusty envelopes shift the emission
of M and C stars toward longer wavelengths. The amount of energy
shifted to longer wavelengths is larger for the young ages, because
more massive and luminous AGB stars are present.

\noindent
It is worth noticing the different IR spectrum of the new SSPs, the
evolution of the features at $11.3\, \mu m$ and $9.7\, \mu m$ in
particular. For young ages, e.g. the SSPs of 0.3 Gyr and 1 Gyr, the
spectrum does not exhibit features due to crystalline silicates,
because the C-stars dominate (the $11.3 \mu m$ feature of $SiC$ is
indeed prominent); for intermediate ages, such as 3 Gyr, the O-stars
of low optical depth influence the spectrum and the $9.7\, \mu m$
feature can be seen in emission. For older ages, from 5 Gyr onward,
the O-stars dominate, so the spectrum becomes more articulated and the
features due to crystalline silicates start to appear at long
wavelengths in the IR.

In Fig.~\ref{SiCSIO} we plot the detailed evolution of the $SiC$ and
$Si-O$ stretching mode features at increasing age. At young ages (0.3,
0.5 Gyr) there is only the feature at $11.3\,\mu m $ of the $SiC$; at
about at 1 Gyr, the $9.7\, \mu m $ of $Si-O$ starts to appear; in the
age range 1 to 2 Gyr the two features overlap; finally, at older ages
the $11.3\, \mu m$ of $SiC$ disappears, and only the feature at $9.7\,
\mu m$ of $Si-O$ occurs.

\begin{figure}
\resizebox{\hsize}{!}{\includegraphics{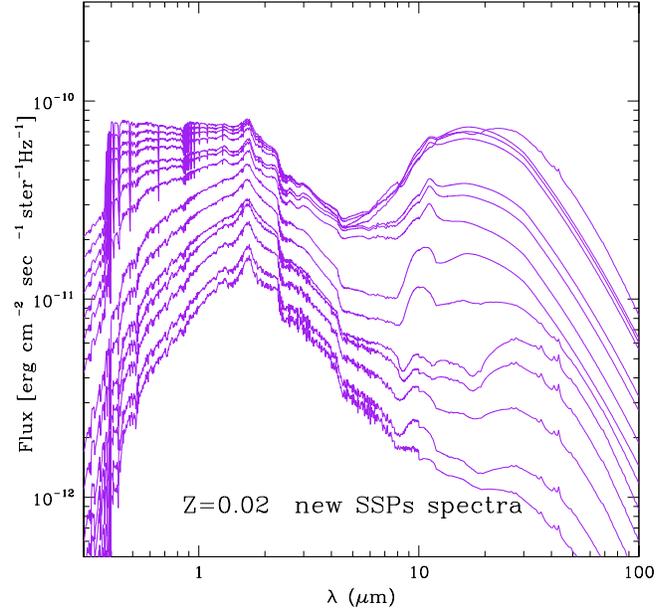}}
\caption{The same as in Fig.~\ref{FluxSSPZ008} but for new SSPs with $Z=0.02$.}
\label{FluxSSPZ02}
\end{figure}

In order to illustrate the effect of the metallicity, we plot in
Fig.~\ref{FluxSSPZ02} the SEDs of the case with $Z$=0.02 for the
same ages of the case with $Z$=0.008 shown in
Fig.~\ref{FluxSSPZ008}.  The evolution of the spectrum is similar
but we can observe an interesting change for the youngest ages
where the $9.7\, \mu m$ feature is slightly in absorption rather
than in emission as at older ages.  The explanation of it can be
seen in Fig.~\ref{AGBz02OC}. For the youngest ages, the luminosity
interval in which C-stars appear becomes very thin, and the stars
at the AGB tip, with the highest mass-loss rate and the highest
optical depth, are O-stars.  So the SSP spectrum becomes dominated
by the $9.7\mu m$ feature in absorption and not in emission, as
expected in envelopes with high optical depth. A similar situation
would also occur at lower metallicities (e.g. $Z=0.008$), but only
at ages younger than 0.2 Gyr as shown by the luminosity intervals
for C-stars displayed in Fig.~\ref{AGBZ008OC}.

\begin{figure}
\resizebox{\hsize}{!}{\includegraphics{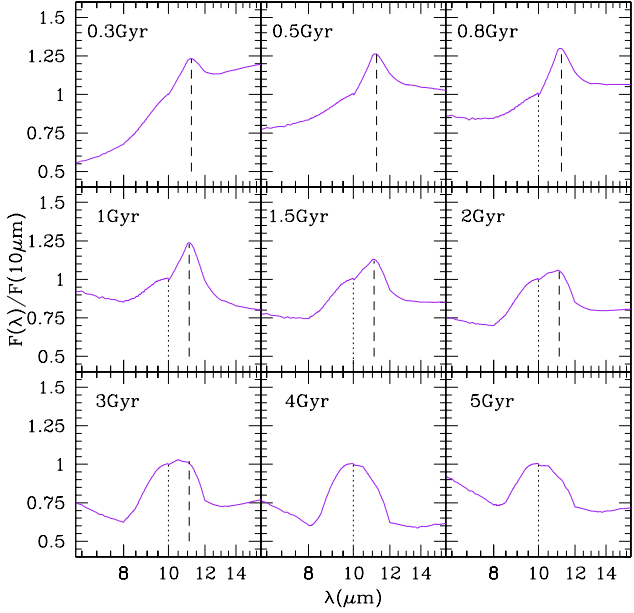}}
\caption{Evolution of the $11.3\, \mu m$ feature of $SiC$  (dashed
vertical lines) and the $9.7\, \mu m$ feature of stretching vibrations
of $Si-O$ bonds (dotted vertical lines) caused by the stretching modes
of this molecule in SSPs of metallicity $Z=0.008$ and ages going from
0.3 Gyr to 5 Gyr. }
\label{SiCSIO}
\end{figure}

\citet{Sil98} calculated spectra of SSPs  with AGB dusty shells
for chemical mixtures containing either silicate or graphite
grains. They do not consider the more realistic situation in which
both types of grain are present and do not include the transition of
O-stars to C-stars (followed in some circumstances by a late
transition from C-stars back to O-stars for AGB stars of very high
mass). Thanks to the inclusion of the transition luminosities of
\citet{Mar99}, it is now possible to follow the IR-SED at changing the
proportions of C-stars with respect to O-stars, which is expected to
vary with the evolution of the AGB star population.

\section{Broad-band colors in the infrared}\label{par:IRcolors}

To test how the presence of dusty shells around AGB stars would affect
the SEDs and colors of individual objects, we derive IRAS and Johnson
broad-band colors and compare them with the observational data for a
selected sample of AGB stars of different type. In addition to this we
calculate the integrated colors of SSPs in the Johnson system and
compare them with a sample of star clusters of the Magellanic Clouds.

{\bf IRAS Colors}. To derive the theoretical IRAS monochromatic fluxes
we have convolved the SEDs of our SSPs with IRAS transmission curves,
derived from the IRAS Explanatory Manual Supplement (on line at
\textit{http://space.gsfc.nasa.gov/astro/iras/docs/exp.sup/}).
Following \citet{Bed87}, the monochromatic flux is defined as

\begin{equation}
F_{\lambda} = \frac{\int F_{\lambda} \Phi_{\lambda} d\lambda}{\int
\left( \frac{F_{\lambda}^{a}}{F_{0\lambda}^{a}} \right)
\Phi_{\lambda} d\lambda}
\label{Bedijn}
\end{equation}

\noindent
where $\Phi_{\lambda}$ is the instrumental profile, $F_{\lambda}$ is
the theoretical flux distribution, $F_{\lambda}^{a}$ is the assumed
flux distribution, that for IRAS bands is $\propto \lambda^{-1}$, and
$F_{0\lambda}^{a}$ is the assumed flux distribution referred to the
central wavelength of the band as defined in \citet{Neu84}. If the
source has a profile with a shape different from the dependence
$\propto \lambda^{-1}$, one has to correct the flux using the
so-called $K$ factor given by

\begin{equation}
K = \frac{\int \left( \frac{F_{\nu}}{F_{\nu_{0}}}
\right)\Phi_{\nu}d\nu}{\int \left(
\frac{F_{\nu}^{a}}{F_{\nu_{0}}^{a}} \right)\Phi_{\nu}d\nu}
\label{kappa}
\end{equation}

\noindent
and to divide the monochromatic flux resulting from eq. (\ref{Bedijn})
by this factor. Finally, the IRAS colors are given by

\begin{equation}
\left[12-25 \right] = -2.5\log \left(\frac{F\left(12\right)}
{F\left(12 \right)_{0}} \right)+2.5 \log \left( \frac{F \left(25
\right)} {F\left( 25 \right)_{0}} \right)
\label{IRASmag}
\end{equation}

\begin{equation}
\left[ 25-60 \right] = -2.5 \log \left(\frac{F\left( 25 \right)}
{F\left(25\right)_{0}}\right)+2.5\log
\left(\frac{F\left(60\right)} {F\left(60\right)_{0}} \right)
\label{IRASmagb}
\end{equation}

\noindent
where $F\left(12\right)$, $F\left(25\right)$, and $F\left(60\right)$
are the IRAS theoretical fluxes in Jansky (Jy), and the constants of
calibration $F\left(12\right)_{0}$=28.3 Jy , $F\left(25\right)_{0}$=6.73
Jy, and $F\left(60\right)_{0}$=1.19 Jy are taken from the IRAS PSC
Explanatory Manual Supplement (1988)
\footnote{We have also calculated the flux $F\left(100\right)$
adopting the calibration constant $F\left(100\right)_{0}$=0.43 Jy.  No
use of this quantity is made here.}.

\begin{figure}
\resizebox{\hsize}{!}{\includegraphics{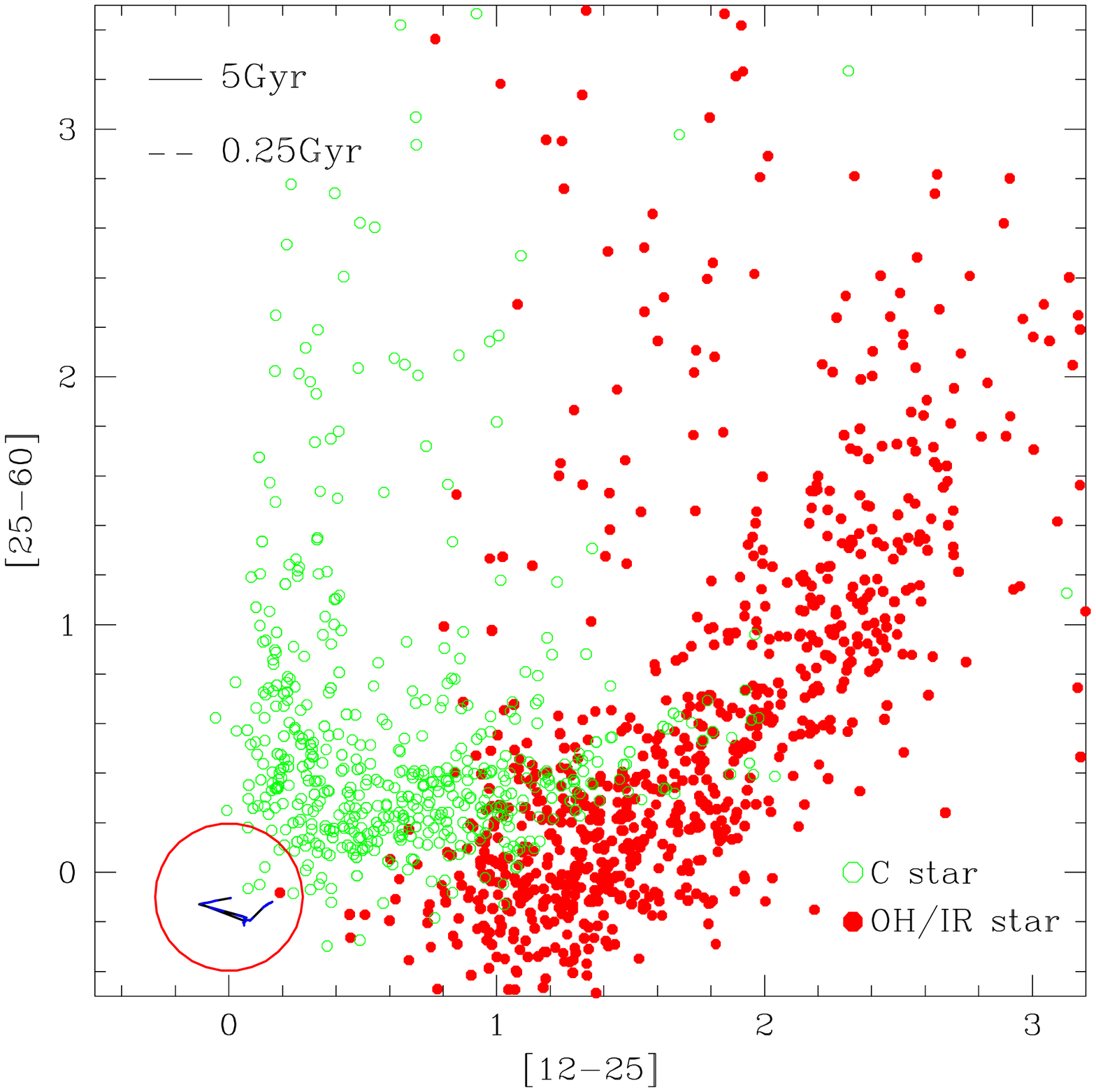}}
\caption{The IRAS two color diagram  $\left[ 25-60 \right]$ vs
$\left[15-25 \right]$. The O-stars have been sampled from different
sources: \citet{Lew90}, \citet{Les92}, \citet{Dav93},
\citet{Blo93}, \citet{Che93}, \citet{Lou93}, \citet{Xio94},
\citet{Lep95}, \citet{Lew97}, \citet{Van98}. The sample of C-stars
is from \citet{Epc90}, \citet{Ega91}, \citet{Cha93}, \citet{Gug93},
\citet{Vol92}, \citet{Gro95}, \citet{Gug98}, \citet{Gug97}.
The fluxes of the stars in the samples are obtained using the IRAS
Point Source Catalogue. Two groups of coeval AGB stars of different
mass represented by the SSPs with age of 0.25 Gyr (dashed line) and 5
Gyr (solid line) are compared to the data. The large circle marks the
color range spanned by these AGB stars. Classical models of AGB stars
are not able to fit the observational color distribution.}
\label{BOTHZ008old}
\end{figure}

Our template sample of O-rich AGB stars (Miras, Semi-Regular, Long
Period Variables, OH/IR etc, indicated by the filled circles) and
C-stars (open circles) is shown in the IRAS two color plane
[25-60] vs [12-25] of Fig.~\ref{BOTHZ008old}.  The location of the
stars in this diagram is widely used to establish the nature of
the sources. \citet{Van88} have defined a region named the
VH-window occupied by late-type stars.  In addition to this, many
other diagnostic plots are used to discriminate between O stars
and C stars. The two-colors plots are constructed combining in
various ways IRAS fluxes with other IR pass-bands, see for
instance \citet{Epc87}. In this diagram we also plot the colors of
the AGB stars.  In order to show the whole range of colors spanned
by these stars we make use of the SSPs: more precisely the colors
displayed by an SSP would correspond to AGB stars of the same age
but different initial mass. By varying the age of the SSPs we may
cover the whole range of colors of AGB stars of any initial mass
and age.

In Fig.\ref{BOTHZ008old} models of classical AGB stars, i.e. with
no dusty shells around, would span a small range of colors
approximately centered at about around $\left[12-25
\right]\backsimeq 0$ and $\left[25-60 \right]\backsimeq 0$ (the
encircled area visualizes the region covered by two SSPs with age
of 0.25 and 5 Gyr). As already pointed out by \citet{Sil98}, this
happens because the SED of AGB stars, whose $T_{eff}$ falls in the
range 2500 to 3500 $K$, is much similar to the Rayleigh-Jeans
distribution of black-body. The net result is that classical
models of AGB stars fail to reproduce the observational
distribution of O and C-stars in the VH-window.

\begin{figure}
\resizebox{\hsize}{!}{\includegraphics{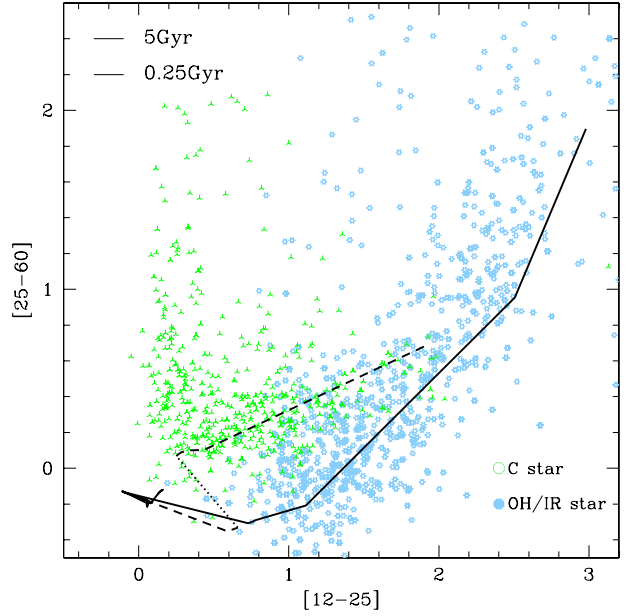}}
\caption{The same as in Fig.~\ref{BOTHZ008old} but in which two
groups of new AGB stars of the same age and different initial mass are
displayed. Like in Fig.~\ref{BOTHZ008old} these AGB stars are
simulated by two SSPs with age of 0.25 Gyr (dashed line) and 5 Gyr
(solid line). Along the line for the young age we mark with a dotted
line the rapid transition from O - to C-stars.}
\label{BOTHZ008r}
\end{figure}

\begin{figure}
\resizebox{\hsize}{!}{\includegraphics{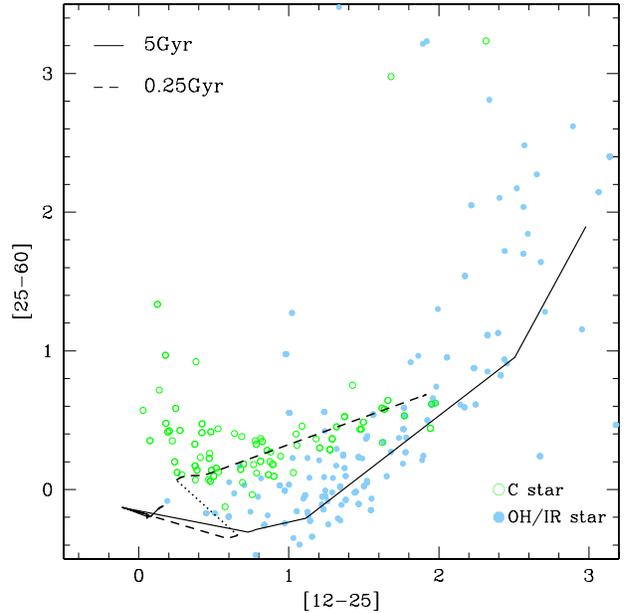}}
\caption{The same as in Figs.~\ref{BOTHZ008old} and
\ref{BOTHZ008r} but in which the data have been selected imposing
the condition $\frac{cirr3}{F_{60}} \leq 3$. All the stars with ratio
$\frac{cirr3}{F_{60}}$ greater than 3 have been discarded because too
much contaminated by the cirrus light.}
\label{BOTHZ008}
\end{figure}

Passing now to the new models of AGB stars with the dusty shell
around, the situation is much improved. This is shown in
Fig.~\ref{BOTHZ008r}, where now the colors of the AGB stars
(represented by two SSPs of given age) stretch across the whole
VH-window.  As we expect, the path in the two color plot of massive
AGB stars (the young SSP of 0.25 Gyr shown by the dashed line) and
low-mass AGB stars (the old SSP of 5 Gyr shown by the solid line) is
different.  The low mass AGB stars overlap only the region occupied by
O-stars, whereas the more massive AGB stars jump into the region
crowed by C-stars when the transition from O- to C-star occurs. This
stage is marked by the dotted portion of the dashed line.

However, it must be noticed that many stars in our sample have
$\left[25-60 \right]$ colors that are significantly redder than
predicted by the theory.

This fact could be easily explained by the so-called cirrus
contamination \citep{Ive95}. In short the cirrus emission
$C_{\lambda}$ may affect the IRAS fluxes because, owing to the
point-like nature of the sources, sky-subtraction may not be accurate
enough and some contamination by the cirrus light can be present and
significant, so that some additional correction of the data is
required The cirrus emission $C_{100}$ can be estimated from IRAS
quantity {\it cirr3}, i.e. the surface brightness at $100\mu m$ around
the point source: $C_{100} = 1.2 \cdot cirr3\, Jy$. \citet{Ive95},
have shown that long-wavelength IRAS point source fluxes are
unreliable when $cirr3 \gtrsim \left(1-5\right)\cdot F_{60}$. Basing
on this, it is possible to clean the sample by removing all the
sources with $\frac{cirr3}{F_{60}} \gtrsim a$, where $1\leq
a\leq5$. In Fig.~\ref{BOTHZ008} are shown the same data of
Fig.~\ref{BOTHZ008r} but in which all stars with
$\frac{cirr3}{F_{60}}\gtrsim 3$ have been removed. Now the theoretical
colors much better reproduce the observed ones. In general at
increasing the cirrus correction, the agreement between theory and
data improves, but too few stars are left over and the comparison
loses statistical significance.

Another point to note is that our models of AGB stars of different
mass do not actually cover the whole color ranges of the data even
considering the cirrus correction above. It is likely that the
parameters adopted in the radiative transfer problem are not fully
adequate.  There are several causes worth being examined in some
detail:

(i) Part of the disagreement can be due to the optical depth of the
envelope and to its dependence on the chemical composition. Recalling
the definition of optical depth, this could vary from model to model
because of different compositions of the dusty envelopes. In this
study we adopt different compositions of the dust for different
intervals of optical depth, however we have not considered the
possibility of different mixtures of silicate and carbon grains in
envelopes with the same optical depth. This is a point of weakness
because \citet{Ive95} have demonstrated that mixed compositions could
explain the spread of the data (see for example their Fig.~6).

(ii) Another possibility, is that the density profile of the matter is
more complicated than a continuous power-law, as a consequence of
discontinuous episodes of mass-loss of different intensity. Therefore
the approximation $\rho \propto r^{-2}$ could be too simple, actually
only mirroring average properties. \citet{Suh97} and \citet{Sil98}
have indeed shown that a narrow region of enhanced density, caused by
a super-wind phase \citep{Ste00} travelling across the envelope, could
be the cause of a large spread in the two color diagram. We plan to
examine this problem in a forthcoming study.

\begin{figure}
\resizebox{\hsize}{!}{\includegraphics{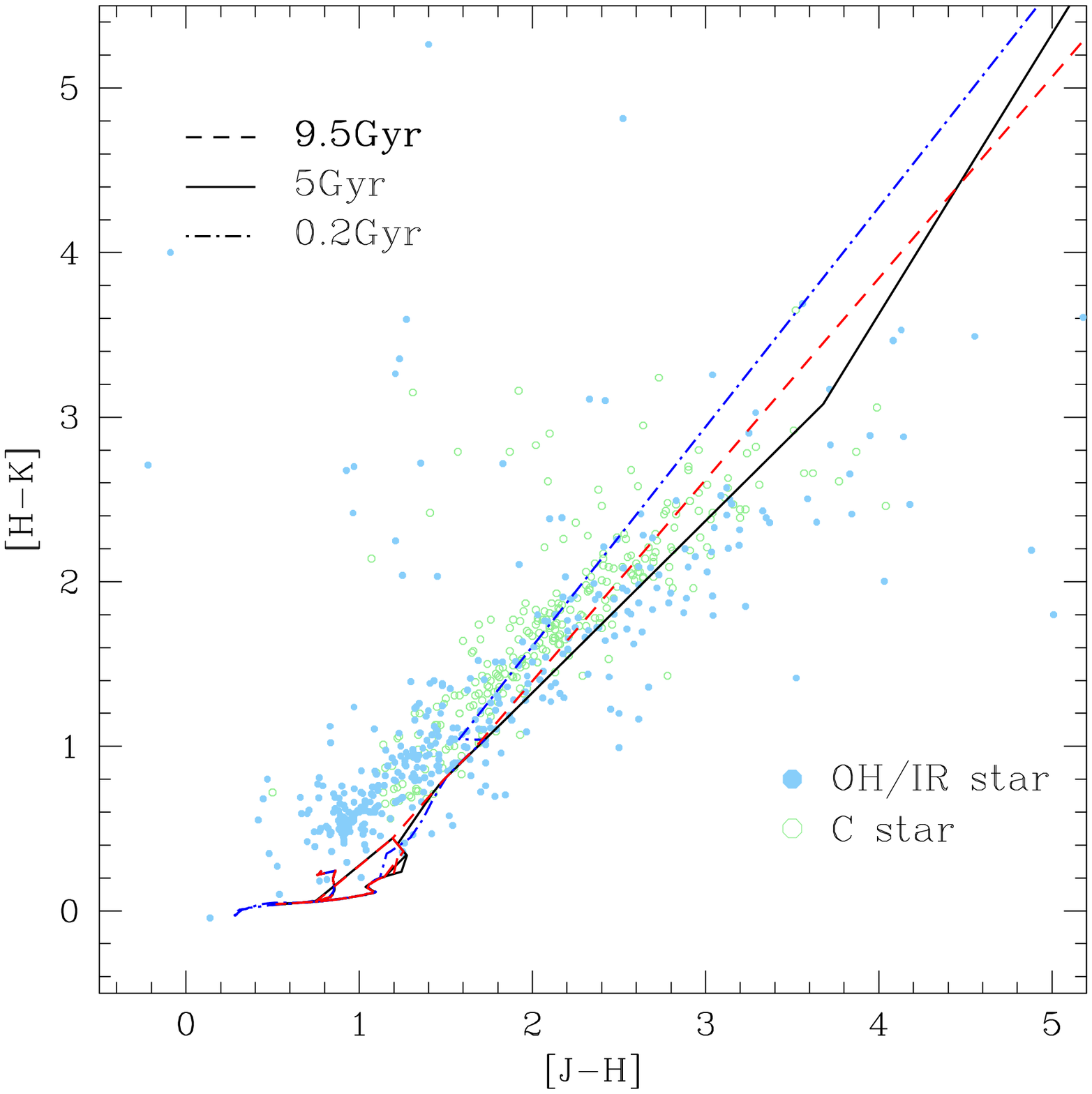}}
\caption{The two color diagram $\left[ H-K \right]$ vs $\left[J-H
\right]$ derived from the new models of AGB stars. The colors of
the OH/IR and Mira stars are taken from \citet{Lep95},
\citet{Xio94}, \citet{Oli01}, \citet{Whi94} whereas those of the
C-stars are derived from \citet{Epc90}, \citet{Gug93},\citet{Gug97},
\citet{Gug98}, \citet{Oli01}. All the data are properly corrected for
extinction. In particular, for the database of \citet{Lep95}, the
infrared data have been corrected with a new model of extinction
(J. R. D. Lepine, private communication). The dotted-dashed, solid and
dashed lines show three groups of AGB stars with age of 0.25, 5 and 9
Gyr respectively.}
\label{JmHHmK}
\end{figure}

\begin{figure}
\resizebox{\hsize}{!}{\includegraphics{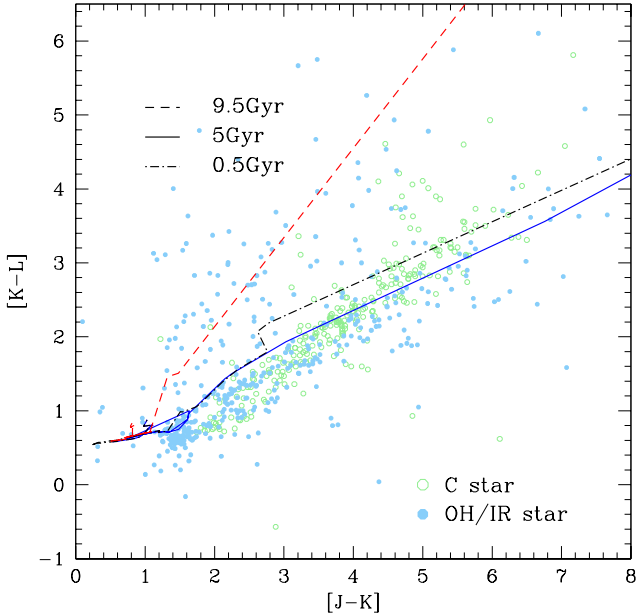}}
\caption{The two color diagram $\left[ K-L \right]$ vs $\left[J-K
\right]$. The sources of data and AGB models are the same as in
Fig.~\ref{JmHHmK}}
\label{JmKKmL}
\end{figure}

{\bf NIR Colors}. We also compare our theoretical colors in near
IR, such as $\left[ J-H \right]$, $\left[ H-K \right]$ and
$\left[K-L \right]$ with the data in the same pass-bands for a
sample of O- and C-stars. The comparison is shown in
Figs.~\ref{JmHHmK} and \ref{JmKKmL}.  Three groups of AGB stars
are shown: massive objects (the SSP of 0.2 Gyr, dotted dashed
line), intermediate mass AGB stars (the SSP of 5 Gyr, solid line),
and low-mass AGB stars (the 9.5 Gyr SSP, dashed line).  While the
classical models for AGB stars are confined in a small region and
thus fail to match the data, the new models with the dusty shells
stretch across the whole region crowded by the observational data.
Both in the $\left[J-H \right]$,$\left[H-K \right]$ and in the
$\left[ J-K \right]$, $\left[K-L \right]$ two color diagrams the
fit is good, even if we still have some problem to fit the data of
O stars. As noticed by \citet{Sil98}, problems in the fit could be
due to the atmospheric models adopted for the M type spectra.

\begin{figure}
\resizebox{\hsize}{!}{\includegraphics{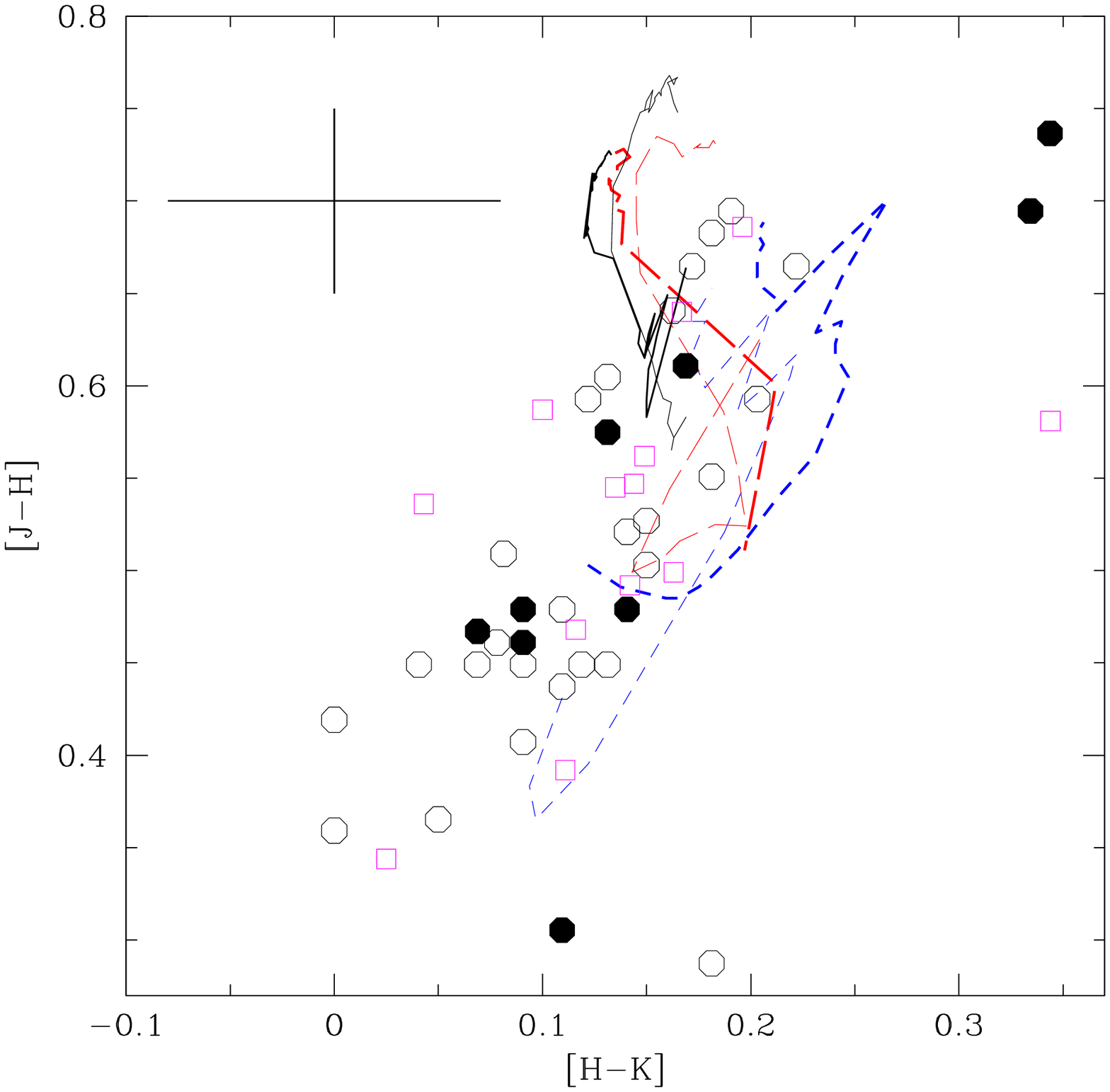}}
\caption{The two color diagram $\left[ J-H \right]$ vs $\left[H-K
\right]$ for young star clusters of the Magellanic Clouds. The
open circles are the LMC clusters selected by \citet{MouI02} from the
catalog of \citet{Per83}, whereas the filled circles are the same but
for SMC clusters. The open squares are a few LMC clusters whose IR
colors have been derived by \citet{Pre02} using the data of the {\it
2MASS Second Incremental Data Release} and companion {\it Image Atlas}
(see the text for more details). All the data have been properly
reddening corrected. The lines show the color range spanned by SSPs of
different metallicity and physical input: the thin and thick dashed
lines are the SSPs of \citet{MouI02} respectively for $Z=0.02$ and
$Z=0.008$; the thin and thick long-dashed lines are our SSPs for
$Z=0.02$ and $Z=0.008$; finally the thin and thick solid lines in the
upper part of the diagram are the old SSPs by \citet{Tantalo98a} for
$Z=0.02$ and $Z=0.008$.  The age of all SSPs is from 100-150 Myr to 15
Gyr.}
\label{graf1}
\end{figure}

\begin{figure}
\resizebox{\hsize}{!}{\includegraphics{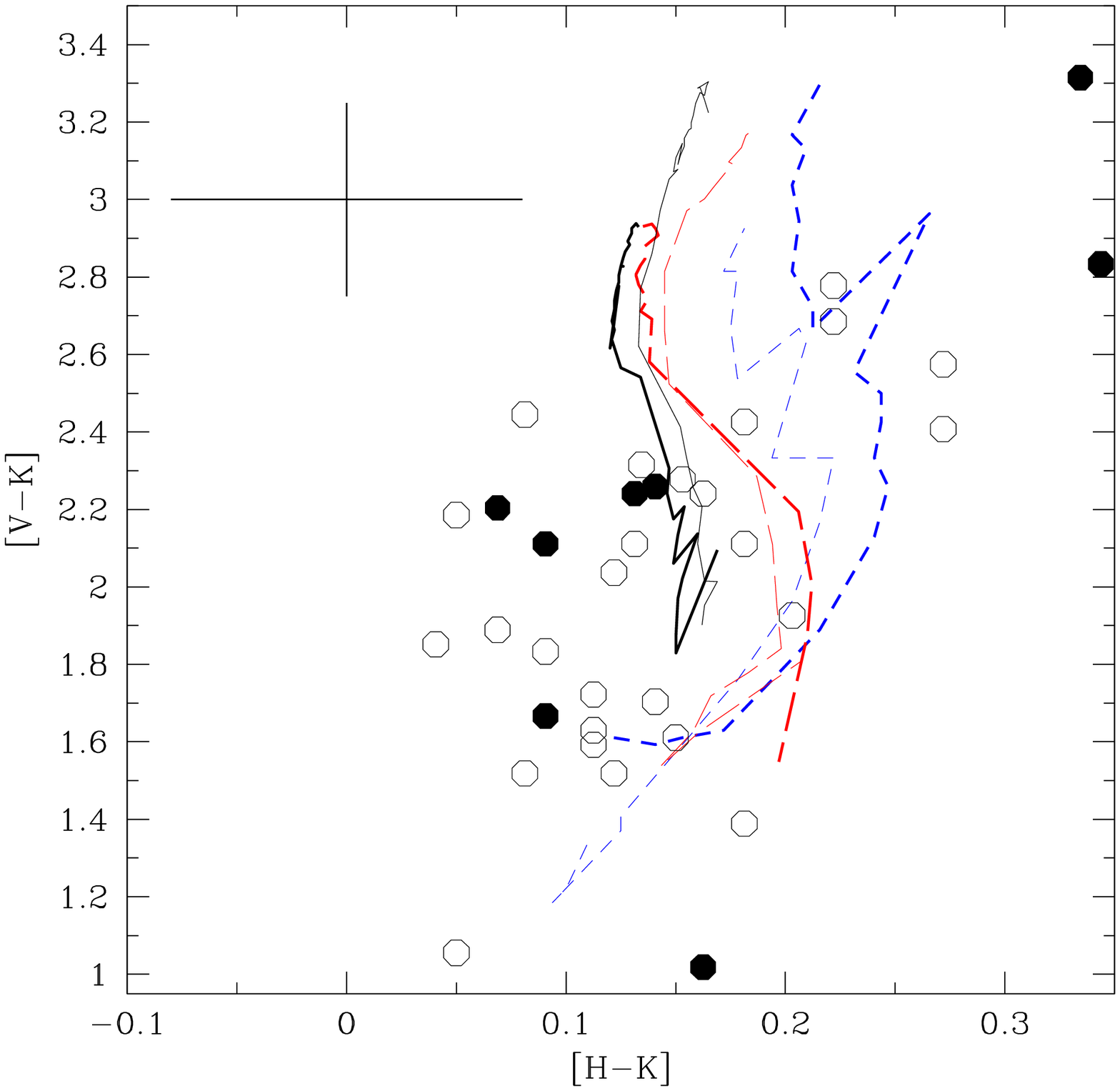}}
\caption{The two color diagram $\left[ V-K \right]$ vs $\left[H-K
\right]$ in the near IR. The meaning of the symbols is the same as
in Fig.~\ref{graf1}. }
\label{graf3}
\end{figure}

\begin{figure}
\resizebox{\hsize}{!}{\includegraphics{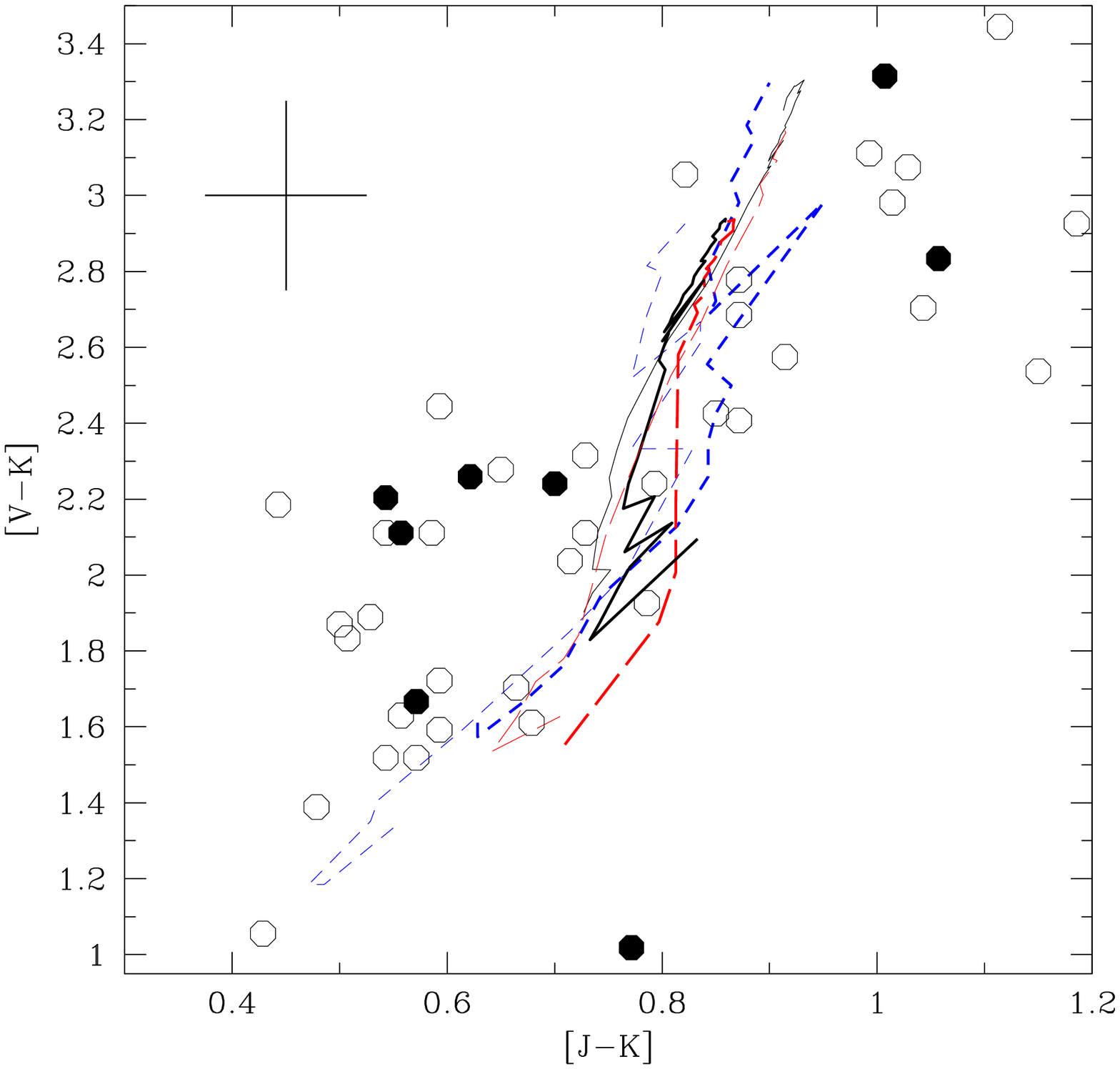}}
\caption{The two color diagram $\left[ V-K \right]$ vs $\left[J-K
\right]$ in the near IR. The meaning of the symbols is the same
as in Fig.~\ref{graf1}.}
\label{graf2}
\end{figure}

\begin{figure}
\resizebox{\hsize}{!}{\includegraphics{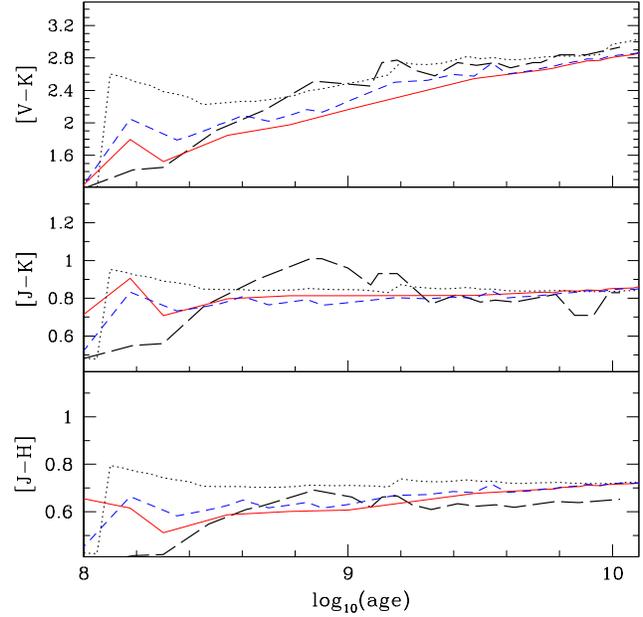}}
\caption{The integrated colors $\left[J-H \right]$, $\left[J-K
\right]$, and and $\left[V-K \right]$ as function of the age in
the range 0.1 to 20 Gyr for SSPs with $Z$=0.008. The dashed lines
are the old SSPs by \citet{Tantalo98a}. The thick solid lines are
the SSPs of this study for the same metallicity. The dotted lines
are the corresponding SSPs by \citet{Gir02}. Finally, the thick
long dashed line shows the SSPs of \citet{MouI02} for $Z=0.008$.}
\label{JmHtime}
\end{figure}

\begin{figure}
\resizebox{\hsize}{!}{\includegraphics{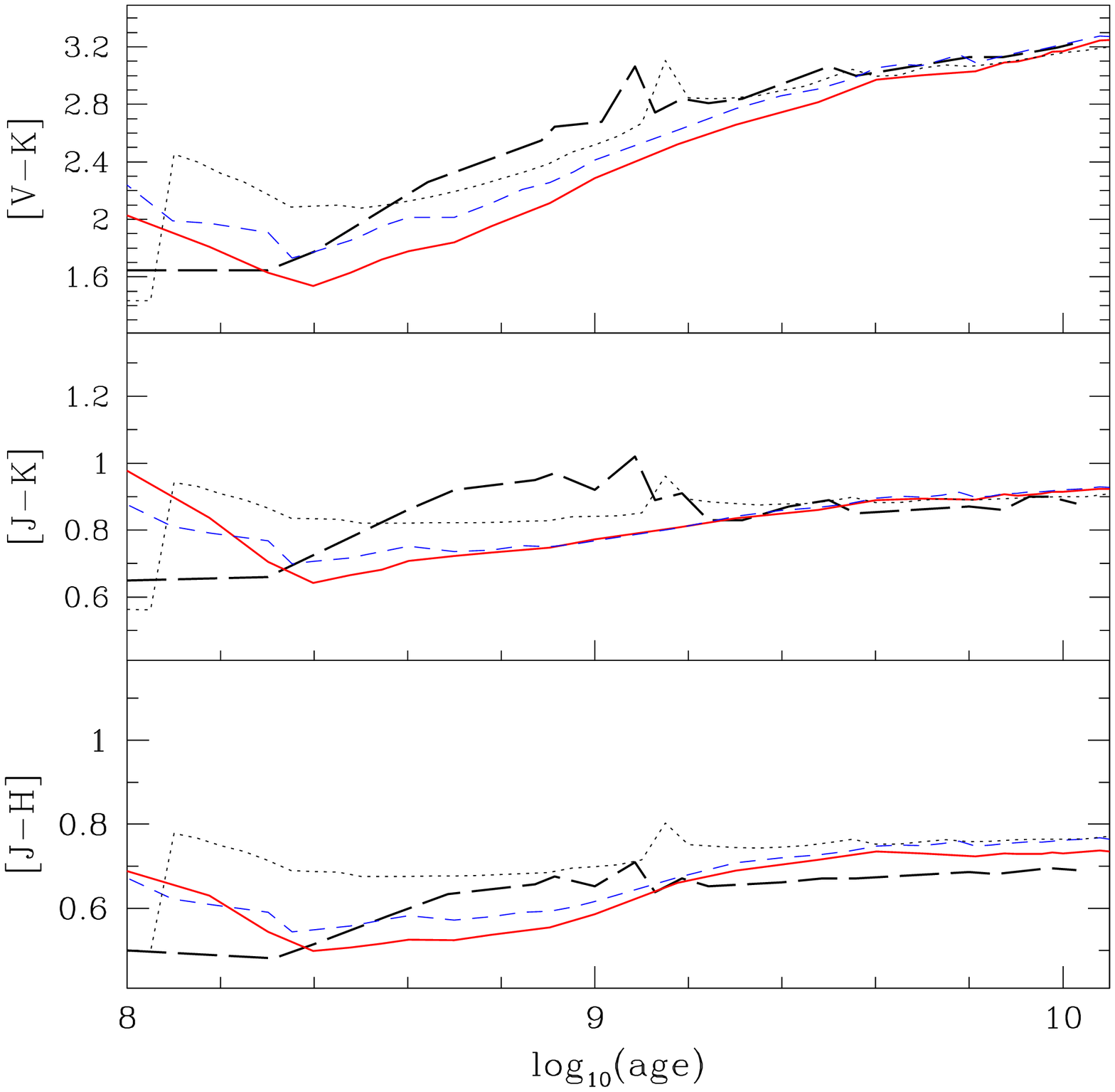}}
\caption{The integrated  color $\left[J-H \right]$, $\left[J-K
\right]$, and and $\left[V-K \right]$ as function of the age in
the range 0.1 to 20 Gyr for SSPs with $Z$=0.02. The meaning of the
symbols is the same of Fig.~\ref{JmHtime}. } \label{JmKtime}
\end{figure}

{\bf Star Clusters}. It might be worth of interest to compare the
integrated broad-band colors of the SSPs whose AGB stars are
enshrouded in the dust shell to those of a sample of star clusters. To
this aim we have looked at the young globular clusters of the Large
and Small Magellanic Clouds (LMC and SMC, respectively). \citet{Pie00}
have presented age determinations for about 600 star clusters
belonging to the central part of the LMC. These clusters are younger
than 1.2 Gyr and therefore the majority of them belong to the age
range in which the AGB phase can develop significantly contributing to
the integrated light of the cluster. Out of the catalog, we have
selected a small sample in which the contribution to the total light
by AGB stars is particularly strong. The photometric data of these
clusters, are derived from the \emph{2MASS Second Incremental Data
Release} and the companion \emph{Image Atlas}, that contains about 1.9
millions of images in $J$, $H$ and $K_{S}$ bands. The integrated
magnitudes $J_i$, $H_{i}$ and $K_{i}$ have been calculated by
\citet{Pre02} and kindly made available to us.  Finally, in order to
compare our SSPs with those by \citet{MouI02}, we consider also the
integrated IR colors for LMC and SMC clusters by
\citet{Per83} used by \citet{MouI02}.

The series of Figs.~\ref{graf1}, \ref{graf3}, and \ref{graf2} show the
planes $\left[J-H\right]$ vs $\left[H-K\right]$, $\left[V-K\right]$ vs
$\left[H-K\right]$, and $\left[V-K\right]$ vs $\left[J-K\right]$,
respectively. In each diagram we display the data of \citet{Per83} for
LMC (open circles) and SMC (filled circles) clusters, the LMC clusters
calculated by \citet{Pre02} (open squares), the SSPs by
\citet{MouI02}, the old SSPs by \citet{Tantalo98a}, and the new ones
of this study. The SSPs of \citet{MouI02} and \citet{Tantalo98a} on
display span an age range from 100-150 Myr (the first ages at which
the AGB contribution is significant) to 15 Gyr and are for the
metallicities $Z=0.008$, and $Z=0.02$. The new SSPs span the same
range of ages and metallicities.

Looking at Figs.~\ref{graf1}, \ref{graf3} and \ref{graf2} we note
that both classical and new SSPs generally agree with the data,
even if in the $\left[J-H \right]$ vs $\left[H-K \right]$ and
$\left[V-K \right]$ vs $\left[H-K \right]$ diagrams old SSPs span
a narrower range in $\left[H-K \right]$ and are too red in
$\left[J-H \right]$. In contrast the new SSPs have bluer
$\left[J-H \right]$ and $\left[V-K \right]$ colors. Although the
agreement is better than with the old SSPs, it is not yet
satisfactory. An additional shift to bluer $\left[J-H \right]$ and
$\left[H-K \right]$ is required. This would imply that the slope
of the SSP SEDs ought to become steeper passing from the $J$ to
the $K$ pass-band than allowed by the present models. Finally the
SSPs by \citet{MouI02} extend toward bluer colors than the other
SSPs. A possible explanation of this unsatisfactory fit is that
the theoretical models of M giants are not yet able to reproduce
the empirical spectra of O and C stars used by \citet{MouII02}. It
is worth recalling that the ultimate reason of the good agreement
achieved by \citet{MouII02} models is the adoption of empirical
spectra. More work is required to improve the theoretical spectra
of O and C-stars to be included in population synthesis studies.

{\bf Temporal evolution}. Finally we look at the temporal
evolution of our theoretical colors. This is shown in
Fig.~\ref{JmHtime} and \ref{JmKtime} for the colors
$\left[J-H\right]$, $\left[J-K \right]$ and $\left[V-K \right]$.
Fig.~\ref{JmHtime} is for the metallicity $Z$=0.008, whereas
Fig.~\ref{JmKtime} is for $Z$=0.020. The age range goes from the
time at which the first AGB stars are formed up to very old ages
when the contribution of AGB stars to the integrated flux gets
very low. For the sake of comparison we show also the colors of
the classical SSPs by \citet{Tantalo98a} and \citet{Gir02} --
these latter are based on more recent stellar models calculations
and stellar libraries -- and finally those of the SSPs by
\citet{MouI02}. The following remarks can be made:

(i) The IR colors of the old SSPs by \citet{Tantalo98a} differ
from those of \citet{Gir02} and also from those of the present
study.

(ii) The IR colors of the new SSPs marginally agree with the
semiempirical ones by \citet{MouI02}. In particular we note that
the IR colors are in good mutual agreement for ages older than
about $1Gyr$, whereas for younger ages, where the contribution of
AGB stars is important, they can significantly differ.

(iii) Finally, it is worth noticing that IR colors from different
sources offer an embarrassing picture. Those of \citet{Gir02} are
redder than any other, except for the $\left[J-K \right]$ bump in
\citet{MouI02} at about $1Gyr$. The colors of \citet{Tantalo98a}
are bluer than those of \citet{Gir02} and corrected for dust
effect surprisingly they get even bluer, with some exceptions. So,
the colors of \citet{Gir02}, \citet{Tantalo98a} and ours form a
sequence from red to blue colors that deserves a careful analysis.
First of all we have to clarify the difference between
\citet{Tantalo98a} and \citet{Gir02} without shells of dust.

\begin{figure}
\resizebox{\hsize}{!}{\includegraphics{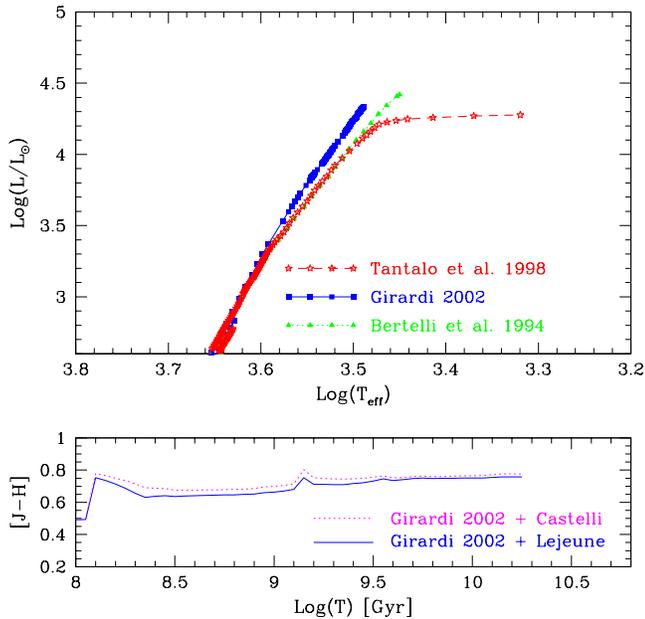}}
\caption{Upper panel: comparison between the AGB of various
isochrones set. We plot the AGB for old \citet{Ber94} isochrones
(filled triangles), \citet{Tantalo98a} isochrones used in this
work (open stars), new \citet{Gir02} isochrones (filled squares).
Lower panel: comparison between [J-H] colors of \citet{Gir02} SSPs
calculated with \citet{Gir02} library of stellar spectra and with
\citet{Lejeune98} theoretical library. } \label{col_agb_tan_gir}
\end{figure}

{\it Spectral libraries.} A plausible cause of disagreement
between \citet{Gir02} IR colors and \citet{Tantalo98a} could be
the different libraries of stellar spectra in usage. \citet{Gir02}
have adopted a library in which some empirical spectra for M giant
stars are included \citep[see][for all details]{Gir02}.
\citet{Tantalo98a} and this study adopt the \citet{Lejeune98}
library in which purely theoretical spectra (even for M-type
stars) are included. To check the effect of different spectral
libraries we have taken the isochrones/SSPs of \citet{Gir02} and
recalculated the IR colors using both the \citet{Gir02} and
\citet{Lejeune98} libraries. The results are compared in the
bottom panel of Fig.~\ref{col_agb_tan_gir} limited to the case of
the [J-H] color. Assigned the set of isochrones/SSPs, no
significant difference arises passing from the library of
\citet{Gir02} to that of \citet{Lejeune98}.

{\it Models of AGB stars.}  Therefore, the cause of disagreement
in the IR colors of Figs.~\ref{JmHtime} and \ref{JmKtime} could be
in the different models for the AGB phase:

(a) {\it Mass-loss.} The first important parameter to look at is
the rate of mass-loss. The SSPs of \citet{Gir02},
\citet{Tantalo98a} and of our study stand on the \citet{Vas93}
prescription (super-wind included), but for the delay correction
proposed by \citet{Vas93} which is not included in \citet{Gir02}.
The effect of mass-loss has therefore little influence on the
colors difference.

(b) {\it Path in the HRD.} Another source of disagreement could be
the distribution of AGB stars in the HR-Diagram. To this aim we
compare in top panel of Fig.~\ref{col_agb_tan_gir} the locus
predicted by three sets of isochrones/SSPs, i.e. \citet{Ber94},
\citet{Tantalo98a} and \citet{Gir02}. The comparison is limited to
the age of 0.3 Gyr and solar metallicity ($Z\simeq 0.02$). At this
age the contribution of the AGB to IR colors is important and the
difference among the sources of IR colors is stronger. We note
that while the \citet{Gir02} and \citet{Ber94} AGB is a straight
line (the one of \citet{Gir02} is even bluer and fainter than
\citet{Ber94}), the AGB of \citet{Tantalo98a} first coincides with
\citet{Ber94} and eventually bends to cooler $T_{eff}$s. All this
can be explained by the slightly different physics adopted in the
underlying stellar models. \citet{Gir02} adopt the same rate of
mass-loss as in \citet{Tantalo98a} but a higher mixing length
parameter compared to \citet{Ber94} and \citet{Tantalo98a}. The
higher mixing length given by \citet{Gir02} is the result of the
lower, more recent opacities, and the calibration of their stellar
models on the Sun \citep[see][for all details]{Gir02}.
\citet{Ber94} and \citet{Tantalo98a} used the same stellar models
and mixing length in turn, but adopted different expressions for
the mass-loss rate and the relationship connecting luminosity,
$T_{eff}$ and core mass (L-$T_{eff}$-Mc). The rate of mass-loss
adopted by \citet{Ber94} is indeed lower than the one used by
\citet{Tantalo98a}. This explains why their AGB phase extends to
bright luminosities. The (L-$T_{eff}$-Mc) relationship adopted by
\citet{Ber94} stand on the previous studies by \citet{Ber90} and
\citet{GroeJong93}, whereas that used by \citet{Tantalo98a}
incorporates the TP-AGB models by \citet{Marigo96}. This accounts
for the different slope in the HR-Diagram of the late AGB phase
passing from \citet{Ber94} to \citet{Tantalo98a}. Incidentally the
(L-$T_{eff}$-Mc) relationship adopted by \citet{Gir02} is much
similar to that of \citet{Ber94}. The different path in the
HR-Diagram bears very much on the final IR colors, because cooler
AGB stars should imply more flux at longer IR wavelengths, so that
a bluer [J-H] color is expected: indeed we have that very cool AGB
stars by \citet{Tantalo98a} yield bluer [J-H] colors, whereas the
hotter AGB stars by \citet{Gir02} yield redder colors.

\begin{figure}
\resizebox{\hsize}{!}{\includegraphics{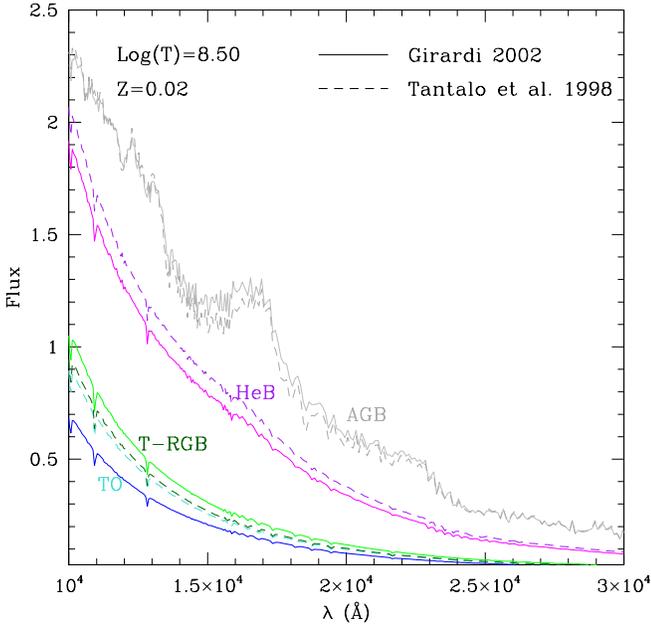}} \caption{The
flux emitted by a SSP with solar metallicity $Z=0.02$ and age of
0.3 Gyr from the stars in different evolutionary stages. We split
the contribution to the total flux into five phases: main sequence
(TO), from TO until RGB tip (T-RGB), horizontal branch (HB),
asymptotic giant branch (AGB) and finally planetary nebulae (PN)
that are not represented for the sake of clarity. The dashed lines
are for the \citet{Tantalo98a} SSP, whereas the solid lines are
same but for the SSP of \citet{Gir02}.} \label{ssp_fluxes}
\end{figure}

(c){\it Number of AGB stars.} Finally, the relative number of AGB
in a SSP must surely play a role. To probe this effect, in
Fig.~\ref{ssp_fluxes} we plot for the SSP with solar metallicity
and age of 0.3 Gyr the cumulative contribution to the total flux
by stars in different evolutionary stages. Five steps are
considered: up to turn-off (TO), from the turn-off up to the tip
of the RGB (T-RGB), from this up to the end of the He-burning
phase (HeB), and from this up the end of the AGB (AGB), and
finally the last phase leading to the formation of Planetary
Nebulae and White Dwarf (this latter is not included in our
analysis owing to the very short lifetime).  Lumping together all
stages up to HeB we note that the cumulative flux predicted by
\citet{Gir02} is lower than that predicted by \citet{Tantalo98a}.
The situation is reversed at the final step when the contribution
from AGB inclusion stars is added: \citet{Gir02} SSPs have higher
fluxes than \citet{Tantalo98a}. This means that a higher
percentage of AGB stars is predicted by \citet{Gir02}. The result
is confirmed by looking at the tabulations of the integral of the
IMF along the isochrones, the so called FLUM defined by
\citet{Ber94} to whom the reader should refer for details. The
reason of it is the kind of (L-$T_{eff}$-Mc) relationship adopted
by \citet{Gir02} which ultimately drives the evolutionary rate
(relative number of stars) during the AGB phase. So, we can
conclude that the combined effect of the path in the HRD and the
relative number of AGB explain the redder colors of \citet{Gir02}.

{\it Effects of dust.}  Finally, there is a fundamental question
to be clarified. Why the inclusion of AGB dusty shells on the
\citet{Tantalo98a} isochrones/SSPs makes their IR colors bluer
than those of \citet{Gir02} instead of making them redder? The
explanation can be found in the effect of the shells of dust on
the stellar radiation of AGB stars. Dust shifts the flux from J,H
and K bands to longer wavelengths, but in general its effect is
stronger at the shorter wavelengths (J band) than at the longer
ones (K band). More flux is driven away from the shorter
wavelengths than the longer ones and this shift tends to increase
the $J$ magnitude more than the $H$ and $K$ magnitudes.

The energy shift changes the flux ratio in a complicated way thus
making it slightly higher or smaller than before. If the flux
ratio is higher the presence of the dusty shells yields the effect
of a redder color: for example we can observe as the [J-H] color
tend to be bluer than one would expect from simple-minded
considerations. For [V-K] color the effect is simpler because dust
increases only the $K$ magnitude without changing the $V$
magnitude and thus producing bluer colors.

\section{Summary, discussion and conclusions}\label{summary}

This paper provides accurate models for the dusty shells surrounding
AGB stars and the re-processing of the radiation emitted by the
central object into the far IR. The aim is to derive realistic SEDs of
AGB stars and SEDs of SSPs to be used in studies of the colors of
individual late-type stars and the integrated colors of stellar
aggregates going from star clusters to galaxies.

The models for the dusty shells take into account the physical
(density and temperature profiles and optical depth) and chemical
structure (dust grains of different compositions as appropriate to
the evolutionary phase under consideration) and solve the
radiative transfer equations to determine the shift of part of the
coming radiation into the far IR region of the spectrum.
Particular attention is payed to the transition of an AGB star
from the O-rich to the C-rich regime taking into account recent
theoretical models of AGB stars.

The SEDs of AGB stars are then convolved with the IRAS and
Johnson-Cousins broad-band pass-bands to derive the IR and far IR
colors to be compared with the observational data for individual
O- and C-stars. The same is made for the SEDs of SSPs to be
compared to the integrated colors of star clusters.  Agreement
between theory and data is good if the effect of the dusty shells
is considered.

Finally, we compare the age dependence of the integrated colors in
the near IR for different sources of SSPs based on different
physical input, i.e. \citet{Tantalo98a}, \citet{Gir02},
\citet{MouI02} and ours. While the various sources more or less
agree for the oldest ages, they differ for the youngest ages, when
there is significative contribution from AGB stars. The
discrepancy can be ascribed to subtle details of the models for
AGB stars proposed by different authors and to the use of
empirical or theoretical libraries for AGB stars. To conclude,
more work is required to derive magnitudes and colors of both
single stars and SSPs in the near and far IR.

Extensive tabulations of spectral energies distributions and
integrated magnitudes and colors for SSPs are available from the
authors upon request.

\begin{acknowledgements}
We would like to thank P. Marigo for very useful discussions and
for providing us the transition luminosities from O-rich to C-rich
AGB stars. We are also grateful to C. Jaeger for making available
to us the absorption coefficients of the forsterite and to J. R.
D. Lepine for correcting data with his new model for the
extinction. Finally, we would like to thank the anonymous referee
for his/her constructive remarks. This study has been financed by
Italian Ministry of Education and Research (MIUR) and the Padua
University with the special contract "Formation and evolution of
elliptical galaxies: the age problem".
\end{acknowledgements}

\bibliographystyle{aa}            
\bibliography{mnemonic,piovan}    

\end{document}